\documentstyle [preprint,revtex,epsf]{aps}
\ifpreprinsty\advance\textwidth by 0.240in\fi

\makeatletter		
\def\footnotesize{\@setsize\footnotesize{9pt}\xpt\@xpt
\abovedisplayskip 8pt plus2pt minus4pt
\belowdisplayskip \abovedisplayskip
\abovedisplayshortskip \z@ plus1pt
\belowdisplayshortskip 4pt plus2pt minus2pt
\def\@listi{\topsep 4pt plus 2pt minus 2pt
\parsep 2pt plus 1pt minus 1pt \itemsep \parsep}}
\def\footnote{\@ifnextchar[{\@xfootnote}{\stepcounter{\@mpfn}%
    \xdef\@thefnmark{\thempfn}\@footnotemark\columnwidth=\hsize\@footnotetext}}
\skip\footins 16pt plus 4pt minus 4pt
\def\fnum@figure{Figure \thefigure.}
\def\figure{\@float{figure}}
\let\endfigure\end@float
\@namedef{figure*}{\@dblfloat{figure}}
\@namedef{endfigure*}{\end@dblfloat}
\makeatother
\begin{document}
\draft
\preprint{UW/PT-92-03}
\preprint{DOE 40614-18}
\preprint{October, 1992}
\begin{title}
Computing Spectral Densities \\
in Finite Temperature Field Theory
\end{title}
\author{Sangyong Jeon}
\begin{instit}
Physics Department FM-15,
University of Washington, Seattle, WA 98105
\end{instit}
\begin{abstract}
Convenient Cutkosky-like
diagrammatic rules
for computing the spectral densities of
arbitrary two-point
correlation functions in finite temperature field theory
are derived.
The approach is based on an explicit analytic continuation of
imaginary-time Feynman diagrams and avoids the complications of real-time
finite temperature perturbation theory. The application of this method
to the perturbative evaluation of transport coefficients is briefly
discussed.
\pagestyle{empty}
\newpage
\end{abstract}

\setcounter{page}{1}

\narrowtext
\section{Introduction}
\label{sec:intro}

Linear response functions characterize the behavior of an equilibrium
thermodynamic system when subjected to a small
disturbance~\cite{Fetter}. Important examples include shear and bulk
viscosities, thermal or electric conductivities, and magnetic
susceptibilities. Any response function may be expressed in terms of a
spectral density defined, for any pair of operators $\hat A$ and
$\hat B$, as
\begin{equation}
\rho_{AB}({\bf k}, \omega)
\equiv
\int  d^3{\bf x}\, dt \, e^{ -i{\bf k} \cdot {\bf x} + i\omega t }
\, \langle [ \hat A ({\bf x},t),\hat B (0)] \rangle
\;.
\label{eq:defsd}
\end{equation}
For example, the static shear viscosity
may be shown to equal~\cite{Hosoya}
\begin{eqnarray}
\eta & \displaystyle = & \displaystyle
i\int d^3{\bf x}\, \int_{-\infty}^0 dt \, \int_{-\infty}^t dt' \,
\, \langle
[
\hat{\pi}_{kl}(0),
\hat{\pi}_{kl}({\bf x},t')
]
\rangle
\nonumber\\
\; & \displaystyle = & \displaystyle
\left.
{1\over 2}
{d \over d\omega}
\rho_{\pi \pi}(0,\omega)
\right|_{\omega=0}
\;,
\label{eq:viscosity}
\end{eqnarray}
where $\hat{\pi}_{kl}({\bf x}, t)$ is
the traceless part of the spatial stress tensor.
For instance, in a scalar field theory
$
\hat{\pi}_{kl}
=
\partial_k \hat{\phi} \, \partial_l \hat{\phi}
-
{\textstyle {1\over 3}}
\delta_{kl}
\partial_m \hat{\phi} \, \partial_m \hat{\phi}
\;.
\label{eq:pikl}
$

Despite the simplicity of the definition (\ref{eq:defsd}), spectral
densities are not simple quantities to evaluate perturbatively at
non-zero temperature. The source of this difficulty is summarized
below. This paper presents a simple reformulation of the standard
diagrammatic rules for a relativistic field theory which will
directly generate the perturbative expansion of a spectral density.

In principle, a spectral density
may be extracted from the corresponding real-time, time-ordered
correlation function,
$
G_{AB}^F({\bf x},t) \equiv
i \langle {\cal T} (\hat{A}({\bf x},t) \hat{B}(0)) \rangle
$.
This correlation function has the
spectral representation~\cite{Fetter},
\begin{eqnarray}
\tilde{G}_{AB}^F ({\bf k}, \omega)
& \displaystyle \equiv & \displaystyle
\int d^3{\bf x}\, dt\,
e^{-i{\bf k} \cdot {\bf x} + i\omega t}
\,
i\langle
{\cal T} (
\hat{A} ({\bf x}, t)
\hat{B} (0)
)
\rangle
\nonumber\\
& \displaystyle = & \displaystyle
\int {d\omega^{\prime} \over 2\pi }
\;
\rho_{AB} ({\bf k}, \omega^{\prime})
\,
\left(
{
1
\over
\omega^{\prime} - \omega -i\epsilon
}
+
2\pi i \, \delta(\omega' - \omega) \, n(\omega')
\right)
\;,
\label{eq:rtcfintermsofsd}
\end{eqnarray}
where $n(\omega') \equiv 1/(e^{\beta\omega'}-1)$ is
the usual Bose-Einstein statistical factor
with inverse temperature $\beta$.
If the operators $\hat A$ and $\hat B$ are Hermitian conjugates,
$\hat{B} = \hat{A}^{\dagger}$,
then as shown in Appendix~\ref{app:cpt},
$\rho_{AB}$ is purely real.
In this case,
the spectral density may easily be
extracted from the real-time correlation function,
\begin{equation}
\rho_{A A^{\dagger}} ({\bf k},\omega)
=
{\rm Im} \;
\tilde{G}_{A A^{\dagger}}^F ({\bf k}, \omega)
\left/
\left(
{\textstyle {1\over 2}}
\coth (\beta\omega /2)
\right)
\right.
\;.
\label{eq:sdfromrtcf}
\end{equation}
Hence, a perturbative expansion of the real-time correlation function
immediately yields the corresponding expansion of the spectral density.

Unfortunately, in contrast to
the situation at zero temperature, at non-zero temperature
the diagrammatic evaluation
of $\tilde{G}_{AB}^F ({\bf k}, \omega)$
requires using an inconvenient
matrix valued propagator. For a scalar field theory,
this has the form~\cite{Niemi},
\begin{eqnarray}
\; &
\displaystyle
{\tilde{G}_0^F} (k)
=
\left(
\begin{array}{cc}
\Delta_{++} (k) \;\; & \displaystyle  \Delta_{+-} (k) \\
\Delta_{-+} (k) \;\; & \displaystyle  \Delta_{--} (k)
\end{array}
\right)
\;,
& \;
\label{eq:rtpropagator}
\\
\noalign{\hbox{where}}
\; & \displaystyle
\Delta_{++}(k) = \Delta_{--} (k)^* =
{i\over k^2 + m^2 -i\epsilon}
-
{2\pi \over e^{\beta |k_0|} - 1}
\delta (k^2 + m^2)
\;,
& \;
\label{eq:delpp}
\\
\noalign{\hbox{and}}
\; & \displaystyle
\Delta_{+-}(k) = \Delta_{-+}(k) =
{
e^{\beta |k_0| /2}
\over
e^{\beta |k_0|} - 1
} \,
2\pi \delta (k^2 + m^2)
\;.
& \;
\label{eq:delpm}
\end{eqnarray}
This structure
arises because the finite temperature,
real-time correlation function
necessarily involves time evolution
operators propagating both forward and backward in time.
To see this, note that
in the interaction picture,
\begin{eqnarray}
& &
\langle
{\cal T} (\hat{A} ({\bf x}, t) \hat{B} (0))
\rangle
\nonumber\\
& & \qquad
=
\lim_{T \to \infty}
{\rm Tr} \left[
\hat{U}_I (-i\beta{-}T, T)
\;
{\cal T} \!
\left(
\hat{A}_I ({\bf x},t)
\hat{B}_I (0)
\hat{U}_I (T, -T)
\right) \,
e^{-\beta\hat{H}_0}
\right]
\biggl/ {\rm Tr}  \;  e^{-\beta\hat{H}}
\;,
\label{eq:ipicturertcf}
\end{eqnarray}
where the subscript $I$ signifies the interaction picture operators,
$\hat{H}$ is the full Hamiltonian and $\hat{H}_0$ is the free
Hamiltonian.
To evaluate this perturbatively,
one must
expand both time evolution operators.
The time evolution operator
that propagates forward
(from $-\infty$ to $+\infty$)
generates
ordinary time-ordered products,
but
the time evolution operator that propagates backward
(from $+\infty$ to $-\infty{-}i\beta$)
generates
anti-time-ordered products.
This difference
cannot be avoided at finite temperature
and is ultimately reflected in the different signs of $i\epsilon$
in $\Delta_{++}$ and $\Delta_{--}$.
Only the
$\Delta_{++}$ part of the propagator
has the right structure to represent real particle propagation.
The other elements of the matrix propagator serve
to cancel unphysical singularities that would occur if only
the $\Delta_{++}$
part of the propagator were used.

This matrix structure substantially increases the difficulty of
evaluating diagrams since each diagram generates many more terms than
at zero temperature.
In addition, one cannot naively
Wick rotate Feynman integrands containing differing $i\epsilon$
prescriptions.

Alternatively, one may obtain the spectral density from the time-ordered
imaginary-time correlation function,
$
G_{AB}^E({\bf x}, \tau) \equiv
\langle {\cal T} (\hat{A}({\bf x},-i\tau) \hat{B}(0)) \rangle
$.
This correlation function is a periodic function of imaginary-time
(with period $\beta$); hence its
Fourier transform depends on
a discrete frequency, $\nu_l=2\pi l/\beta$,
where $l$ is an integer.
The imaginary-time correlation function has the spectral representation,
\begin{eqnarray}
\tilde{G}_{AB}^E ({\bf k}, i\nu_l)
& \displaystyle \equiv & \displaystyle
\int_0^{\beta} d\tau \int d^3{\bf x}\,
e^{-i{\bf k} \cdot {\bf x} + i\nu_l\tau}
\langle
{\cal T}
(
\hat{A} ({\bf x}, -i\tau)
\hat{B} (0)
)
\rangle
\nonumber\\
\, & \displaystyle = & \displaystyle
\int {d\omega \over 2\pi}\,
{
\rho_{AB} ({\bf k}, \omega)
\over
\omega - i\nu_l
}
\;.
\label{eq:itcfintermsofsd}
\end{eqnarray}
To obtain the spectral density,
one must
analytically continue
the frequency dependence of $\tilde{G}_{AB}^E$ from the discrete
set $\{ i\nu_l \}$ to an arbitrary complex value,
and then take the discontinuity across the real axis,
\begin{eqnarray}
i\rho_{AB} ({\bf k}, \omega)
& \displaystyle = & \displaystyle
{\rm Disc} \;
\tilde{G}_{AB}^E ({\bf k}, \omega)
\nonumber\\
\;
& \displaystyle \equiv & \displaystyle
\tilde{G}_{AB}^E ({\bf k}, \omega+i\epsilon)
-
\tilde{G}_{AB}^E ({\bf k}, \omega-i\epsilon)
\;.
\label{eq:sdfromitcf}
\end{eqnarray}
Unlike Eq.~(\ref{eq:sdfromrtcf}), which holds only when the spectral
density is real, the relation (\ref{eq:sdfromitcf}) is valid for any
pair of operators.
If $\hat{B} = \hat{A}^{\dagger}$, so that the spectral density is
real, then the discontinuity (\ref{eq:sdfromitcf}) reduces to
$2\tanh (\beta\omega / 2)$ times the imaginary part of the real-time
time-ordered correlation function.

The perturbative evaluation of finite temperature, imaginary-time
correlation functions involves precisely the same momentum space
diagrammatic rules as at zero temperature except for the replacement
of frequency integrals by sums~\cite{Fetter,Brown,Kapusta},
$\int {dk^0 / 2\pi} \rightarrow
(1 / \beta)\sum_{\nu_l}$.
In contrast to the complicated structure of the real-time
propagator (\ref{eq:rtpropagator}), the free scalar propagator for
the perturbative calculation of imaginary-time correlation functions
is simply the usual
\begin{equation}
\tilde{G}_0^E ({\bf k}, i\nu_l)
=
{1
\over
\nu_l^2 + {\bf k}^2 + m^2
}
\;.
\label{eq:itpropagator}
\end{equation}
For determining the spectral density,
the drawback of the imaginary-time approach is that
one must first manage to perform explicitly
all the frequency summations.
Only then can one analytically continue the external frequency
and compute the discontinuity (\ref{eq:sdfromitcf}).

As shown in the next section,
all the necessary frequency summations can be evaluated directly.
Furthermore, one may perform these sums, analytically continue
the external frequency, and take the discontinuity across the real axis
before performing any of the remaining spatial momentum integrals.
This process generates many terms from a single Feynman diagram, with
individual terms corresponding to
each possible ``cutting'' of
an old-fashioned time-ordered diagram.
These terms may be recombined to obtain simple expressions
corresponding to ``cut'' Feynman diagrams which are generated by
a set of rules that
closely resemble the zero temperature Cutkosky rules.

In this paper we consider only two-point correlation functions.
Finite temperature Cutkosky rules for general $n$-point functions
are discussed in reference~\cite{Kobes}
using different methods.
In the following, we first review
the calculation of imaginary-time correlation functions,
and show how
the process of analytic continuation and computation of the discontinuity
may be incorporated into the diagrammatic rules.
For non-relativistic theories, discussion of a portion of this
approach may be found in reference~\cite{Baym}.
The resulting method is then applied to a simple example of the
spectral density of the composite operator correlation function
$i\langle {\cal T}
(\hat{\phi}^2 ({\bf x}, t) \hat{\phi}^2 (0) )
\rangle$.
Particular attention is focused on the low frequency behavior, as this is
the limit relevant to the calculation of transport coefficients.
Appendix~\ref{app:cpt}
discuss the symmetry properties of the spectral density and the
implications of $C\!PT$ invariance.
For ease of the presentation,
the body of this paper discusses theories
with zero chemical potential and composite operators without time
derivatives.
Appendix~\ref{app:derivatives} discusses
the treatment of operators involving time
derivatives of the elementary field,
and the generalization to non-zero chemical potential is presented in
Appendix~\ref{app:chemicalpotential}.

\narrowtext
\section{finite temperature perturbation theory}
\label{sec:ftpt}

Consider, for simplicity,
a relativistic scalar field theory at non-zero temperature
with vanishing chemical potential.
By applying the standard momentum space Feynman rules,
the perturbation series
for a Fourier transformed imaginary-time
correlation function may be generated.
However, as the number of the loops increases,
dealing with the discrete frequency summations becomes inconvenient.
To avoid these frequency sums (at the cost of further integrals)
one may simply Fourier transform (in time) the momentum space
propagators. This yields
``mixed'' propagators that depend on a spatial momentum
and imaginary-time~\cite{Pisarski}.
To evaluate this mixed propagator for a scalar theory,
it is convenient to first write the momentum space propagator in terms of
the free particle spectral density $\rho_0 ({\bf k}, \omega,)$,
\begin{eqnarray}
\; &
\displaystyle
\tilde{G}_0^{E} ({\bf k}, i\nu_l)
=
{1
\over
\nu_l^2 + {\bf k}^2 + m^2
}
=
\int {d\omega \over 2\pi}\,
{
\rho_0 ({\bf k}, \omega)
\over
\omega - i\nu_l
}
\;,
& \;
\label{eq:propagatorintermsofsd}
\\
\noalign{\hbox{where}}
\; &
\displaystyle
\rho_0 ({\bf k},\omega) =
{2\pi\over 2 E_k}
\Bigl(
\delta(\omega - E_k) - \delta(\omega + E_k)
\Bigr)
\;,
& \;
\label{eq:freesd}
\end{eqnarray}
and $E_k = \sqrt{ {\bf k}^2 {+} m^2 }$.
Then, using the relations
\begin{equation}
{1 \over \beta}
\sum_{l=-\infty}^{\infty} e^{-i\nu_l\tau}
{ 1 \over \omega - i\nu_l }
=
\theta (\tau) {e^{-\omega\tau} \over 1 - e^{-\beta\omega}}
+
\theta (-\tau) {e^{-\omega\tau} \over e^{\beta\omega} - 1}
\label{eq:sumform}
\end{equation}
(valid in the interval $-\beta \leq \tau \leq \beta$),
and
$-n(-\omega) = (1 {+} n(\omega))$
one obtains
\begin{eqnarray}
\tilde{G}_0^E ({\bf k}, \tau)
& \displaystyle = & \displaystyle
\sum_{l=-\infty}^{\infty}
\tilde{G}_0^E ({\bf k}, i\nu_l) \, e^{-i\nu_l\tau}
\nonumber\\
& \displaystyle = & \displaystyle
\int_{-\infty}^{\infty} {d\omega \over 2\pi}\,
\rho_0({\bf k},\omega) \,
e^{-\omega\tau} \,
\Bigl(
\theta (\tau)
( 1 + n(\omega) )
+
\theta(-\tau) n(\omega)
\Bigr)
\nonumber\\
& \displaystyle = & \displaystyle
\int_{-\infty}^{\infty} {d\omega \over 2\pi}\,
\Bigl( 1 + n(\omega) \Bigr) \,
\Bigl(
\rho_0 ({\bf k},\omega)    \,   \theta (\tau)
- \rho_0 ({\bf k},-\omega)  \,   \theta (-\tau)
\Bigr) \,
e^{-\omega|\tau|}
\;.
\label{eq:mixedone}
\end{eqnarray}
For the
scalar field propagator,
the rotational invariance of the system implies that the spectral density
depends only on the magnitude of the momentum.
Since the free spectral density (\ref{eq:freesd})
is an odd function of the frequency,
the propagator takes the following convenient form,
\begin{equation}
\tilde{G}_0^E (|{\bf k}|, \tau)
=
\int_{-\infty}^{\infty} {d\omega \over 2\pi}\,
\Bigl( 1+n(\omega) \Bigr) \,
\rho_0(|{\bf k}|,\omega) \,
e^{ -\omega |\tau|}
\;.
\label{eq:mixedpropagator}
\end{equation}
This result may, of course, be derived directly using the free
equation of
motion plus periodicity in imaginary time.

As discussed in Appendix~\ref{app:cpt},
in the absence of a chemical potential
$C\!PT$ invariance
guarantees that the spectral density for the interacting
scalar field propagator remains an odd function of the frequency.
In what follows,
this symmetry of the spectral density will play a role in simplifying
the diagrammatic rules.

To evaluate a Feynman diagram for a two-point function
using these ``mixed'' propagators, one
labels
each line by a spatial momentum and
each vertex by a Euclidean time $\tau_j$ which will be
integrated from $0$ to $\beta$.
At each vertex where an external
operator inserts external frequency $i\nu_j$,
an additional factor of
$\exp\left(i\nu_j \tau_j \right)$
is present.
Each interaction vertex contributes a factor of $(-\lambda)$.
Each line connecting two vertices labeled by the times
$\tau_a$ and $\tau_b$
represents a propagator
$\tilde{G}_0^E (|{\bf k}|, \tau_a{-}\tau_b)$.
As usual, spatial momenta are conserved at every vertex and all
remaining spatial loop momenta are integrated over.
Thus, the contribution of a diagram
with a total of $n{+}1$ vertices,
two operator insertions
and $L$ loops has the schematic form
\begin{eqnarray}
{C}^{AB}_{\Gamma} ({\bf q}, i\nu)
& \equiv & \displaystyle
{ (-\lambda)^{n-1}
\over
{\cal S}_{\Gamma} }
\int_0^{\beta} d\tau_0 \cdots d\tau_{n} \,
\int {d^3{\bf k}_1 \over (2\pi)^3} \cdots
{d^3{\bf k}_L \over (2\pi)^3} \,
\exp
\Bigl\{
i\nu\tau_l + i\nu'\tau_0  \;
\Bigr\}
f^{AB}( \{ {\bf k}_L \}, {\bf q} )
\nonumber\\
\; & \displaystyle \; & \displaystyle
{}
\times
\prod_{\alpha \in \Gamma }
\int {dk^0_{\alpha} \over 2\pi} \,
\Bigl( 1+n(k^0_{\alpha}) \Bigr) \,
\rho_0 (|{\bf k_{\alpha}}|, k^0_{\alpha}) \,
\exp
\Bigl\{
-k^0_{\alpha} |\tau_a - \tau_b|
\Bigr\}
\;.
\label{eq:schematicint}
\end{eqnarray}
Here, $\Gamma$ denotes the Feynman diagram under consideration,
$\alpha$ labels the different lines of the graph, and
${\cal S}_{\Gamma}$
is the overall symmetry factor.
The factor
\begin{equation}
f^{AB}( \{ {\bf k}_L \}, {\bf q} )
\equiv
f^{A}( \{ {\bf k}_L \}, {\bf q} ) \,
f^{B}( \{ {\bf k}_L \}, {\bf q} )
\;
\end{equation}
represents the action of the external operators $\hat{A}$ and
$\hat{B}$. It is only a function of spatial momenta
provided that the operators
do not contain any time derivatives of the elementary field.
If they do contain time derivatives,
then so does
$f^{AB}( \{ {\bf k}_L \}, {\bf q} )$.
For simplicity, we will assume that no such time derivatives are
present.
The generalization to include
operators with time derivatives is discussed in
Appendix~\ref{app:derivatives}; this
presents minor complications but introduces no new features.
The times $\tau_l$ and $\tau_0$ label the vertices where the external
operators insert the external frequencies $i\nu$ and $i\nu'$
and $\{ \tau_a, \tau_b \}$ are times within the set
$\{ \tau_0, \cdots, \tau_n \}$.

Due to the translational invariance in imaginary time,
one can immediately do one time
integration which generates an
overall Kronecker delta function
enforcing the conservation of the total external frequency,
$i\nu {+} i\nu' = 0$.
Equivalently, one may simply set $\tau_0$ to zero
and assign $i\nu' = -i\nu$ to that vertex.
Then, the remaining time integrals in ${C}^{AB}_{\Gamma}$
have the form
\begin{equation}
{\cal J} =
\int_0^{\beta} d\tau_{1} \cdots d\tau_{n} \,
\exp
\Bigl\{
i\nu \tau_l  \;
- \!
\sum_{\alpha} \,
k^0_{\alpha} |\tau_a - \tau_b|
\Bigr\}
\;,
\label{eq:reducedschematic}
\end{equation}
with $\tau_0 \equiv 0$.
The evaluation of this integral and the association of the result with
old-fashioned time-ordered diagrams
are the main subjects of this section.
Performing these intermediate time integrals in
this ``mixed'' form of perturbation
theory is exactly equivalent to carrying out
all the intermediate frequency summations in the Fourier
transformed version of perturbation theory.
After performing these integrations,
it will be easy to analytically continue
the external frequency in the resulting expression.

Before proceeding with the general analysis, it may be helpful to
consider a simple example.
The lowest order connected term in the perturbative expansion of
$\langle
{\cal T}
(
\hat{\phi}^2 ({\bf x}, -i\tau)
\hat{\phi}^2 (0)
)
\rangle$
is given by twice
the square of the free propagator,
$(G_0^E ({\bf x}, \tau))^2$.
The Fourier transform is
\begin{eqnarray}
\tilde{G}_{\phi^2 \phi^2}^{\hbox{\scriptsize{1-loop}}} ({\bf q}, i\nu)
\; & \equiv & \displaystyle
2
\int d^3 {\bf x}
\int_0^{\beta} d\tau \,
e^{-i{\bf q}\cdot{\bf x} + i\nu\tau} \,
\Bigl(
G_0^E ({\bf x}, \tau)
\Bigl)^2
\nonumber\\
\; & = & \displaystyle
2
\int {d^3 {\bf k} \over (2\pi)^3}
\int_0^{\beta} d\tau \,
e^{i\nu\tau} \,
\tilde{G}_0^E (|{\bf k}|, \tau)
\tilde{G}_0^E (|{\bf q} - {\bf k}|, \tau)
\nonumber\\
\; & = & \displaystyle
2
\int {d^3 {\bf k} \over (2\pi)^3}
\int_0^{\beta} d\tau \,
e^{i\nu\tau}
\int
{dk^0 \over 2\pi}\,
e^{ -k^0 |\tau|}
\Bigl( 1+n(k^0) \Bigr) \,
\rho_0(|{\bf k}|,k^0) \,
\nonumber\\
\; & \; & \displaystyle
\qquad\qquad\qquad\qquad
\times
\int
{dk^{\prime 0} \over 2\pi}\,
e^{-k^{\prime 0} |\tau|}
\Bigl( 1+n(k^{\prime 0}) \Bigr) \,
\rho_0(|{\bf q}-{\bf k}|,k^{\prime 0}) \,
\nonumber\\
\; & = & \displaystyle
2
\int {d^3 {\bf k} \over (2\pi)^3}
\int
{dk^0 \over 2\pi}\,
{dk^{\prime 0} \over 2\pi}\,
\Bigl( 1+n(k^0) \Bigr) \,
\rho_0(|{\bf k}|,k^0) \,
\nonumber\\
\; & \; & \displaystyle
\qquad \times
\Bigl( 1+n(k^{\prime 0}) \Bigr) \,
\rho_0(|{\bf q} - {\bf k}|,k^{\prime 0}) \,
{
    1 - e^{-\beta(k^0 + k^{\prime 0})}
        \over
    k^0 + k^{\prime 0} - i\nu
}
\;.
\label{eq:lowestterm}
\end{eqnarray}
To simplify this, we may use
$e^{-\beta k^0}\,
( 1 {+} n(k^0) )
=
-
( 1 {+} n(-k^0) )$,
plus
the fact that the spectral density of
the propagator is an odd function of the frequency,
to rewrite the result as
\begin{eqnarray}
\tilde{G}_{\phi^2 \phi^2}^{\hbox{\scriptsize{1-loop}}} ({\bf q}, i\nu)
& = & \displaystyle
2
\int {d^3 {\bf k} \over (2\pi)^3}
\int
{dk^0 \over 2\pi}\,
{dk^{\prime 0} \over 2\pi}\,
\Bigl( 1+n(k^0) \Bigr) \,
\rho_0(|{\bf k}|,k^0)
\Bigl( 1+n(k^{\prime 0}) \Bigr) \,
\rho_0(|{\bf q} - {\bf k}|,k^{\prime 0})
\nonumber\\
\qquad & \; & \displaystyle
\qquad \times
\left(
{
          1
        \over
    k^0 + k^{\prime 0} - i\nu
}
+
{
          1
        \over
    k^0 + k^{\prime 0} + i\nu
}
\right)
\;.
\label{eq:reducedlowest}
\end{eqnarray}

Aside from the fact
that the external frequency is discrete and imaginary,
the terms in the parenthesis of Eq.~(\ref{eq:reducedlowest}) may be
recognized as the standard energy denominators of time-dependent
perturbation theory \cite{Ziman}.
The integration over the spatial momentum
${\bf k}$ can be
regarded as the sum over all intermediate two-particle states.
Thus it is natural to associate
the two terms in Eq.~(\ref{eq:reducedlowest})
with old-fashioned time-ordered perturbation theory diagrams
as illustrated in Figure~\ref{fig:lowest}.
\begin{figure}
\begin{picture}(100, 100)(10,20)
\put (180,152) {${\bf q} {-} {\bf k}$}
\put (151,190) {$( {\bf q}, i\nu )$}
\put (151,28) {$( {\bf q}, i\nu )$}
\put (108,60) {${\bf k}$}
\put (332,152) {${\bf q} {-} {\bf k}$}
\put (303,190) {$( {\bf q}, i\nu )$}
\put (303,28) {$( {\bf q}, i\nu )$}
\put (260,60) {${\bf k}$}
\put (50,20) {\vector(0,1){180}}
\put (20, 110) {time}
\end{picture}
\leavevmode\epsfbox{lowest.ps}
\caption{%
\advance\baselineskip by -8pt
The lowest order diagram for the spectral density of
$\langle {\cal T} ({\hat{\phi}}^{2} ({\bf x}, t)
\hat{\phi}^2 (0)) \rangle$.
}
\label{fig:lowest}
\end{figure}
The time ordering of the first diagram corresponds to
the energy denominator $(k^0 {+} k^{\prime 0} {-} i\nu)$ and
the time ordering of the second diagram corresponds to
the energy denominator $(k^0 {+} k^{\prime 0} {+} i\nu)$.
In both cases, the energy denominator can be interpreted as the sum of
frequencies of lines crossing the interval between two external
vertices minus the external frequency flowing out of the vertex above
the interval.
The contribution of a line with momentum
$({\bf k}, k^0)$ is given by
$( 1 {+} n(k^0) )\, \rho_0 ( |{\bf k}|, k^0 )$, which in the zero
temperature limit reduces to
$\theta(k^0) 2\pi \delta( k^2 {+} m^2 )$.
Integration over the spatial loop momentum and frequencies $k^0$ and
$k^{\prime 0}$ completes the expression.

This association of the result of the imaginary-time integrations
(or equivalently the discrete frequency summation)
with old-fashioned time-ordered diagrams
is a general result. To see this, we now go back to the evaluation
of the time integral in (\ref{eq:reducedschematic}),
$$
{\cal J} =
\int_0^{\beta} d\tau_{1} \cdots d\tau_{n} \,
\exp
\Bigl\{
i\nu \tau_l  \;
- \!
\sum_{\alpha} \,
k^0_{\alpha} |\tau_a - \tau_b|
\Bigr\}
\;,
$$
and split the integrand into $n!$ terms
according to the relative
orderings of the $n$ different time variables.
The term corresponding to a particular time ordering, for example,
$\tau_{n}~\geq~\tau_{n-1}~\geq~\cdots~\geq~\tau_2~\geq~\tau_1$,
may be written as
\begin{equation}
{\cal I} (\{ k^0_{\alpha} \}, i\nu)
\equiv
\int_{0}^{\beta} d\tau_{n} \cdots \int_{0}^{\tau_3} d\tau_2
\int_{0}^{\tau_2} d\tau_1 \,
      \exp \Bigl\{
                 \sum_{s=1}^{n} \,
                     - \sigma_s \tau_s
           \Bigr\}
\;,
\label{eq:onetypicalterm}
\end{equation}
where
\begin{eqnarray}
\sigma_s
\equiv
\omega_{\rm in}^s - \omega_{\rm out}^s - i\nu_s
\;,
\label{eq:sigmal}
\\
\noalign{\hbox{with}}
\omega_{\rm in}^s
\equiv \sum_{
            \scriptstyle
            \alpha \in \{
                        (s) \leftrightarrow
                              \atop
            \scriptstyle
                       (0,1,2,\cdots,s-1)
                       \}
} \!\! k^0_{\alpha}
\;,
\\
\noalign{\hbox{and}}
\omega_{\rm out}^s
\equiv \sum_{
            \scriptstyle
            \alpha \in \{
                        (s) \leftrightarrow
                              \atop
            \scriptstyle
                        (s+1,s+2,\cdots,n)
                       \}
} \!\! k^0_{\alpha}
\;.
\label{eq:omegaout}
\end{eqnarray}
Here,
$\left\{ (a,\cdots,b) \leftrightarrow (c,\cdots,d) \right\}$
indicates the set of lines
connecting vertices at times
$\{ \tau_a,\cdots,\tau_b \}$ with those at times
$\{ \tau_c,\cdots,\tau_d \}$.
Hence,
$\omega_{\rm in}^s$
is the sum of frequencies of all lines connecting $\tau_s$ with
earlier times $\{ \tau_0,\tau_1,\tau_2,\cdots,\tau_{s-1} \}$,
and
$\omega_{\rm out}^s$
is the sum of frequencies of the lines connecting $\tau_s$ with
later times $\{ \tau_{s+1},\tau_{s+2},\cdots,\tau_n \}$.
The discrete external frequency $i\nu_s$ is given by
$i\nu\, \delta_{sl}$ because the external frequency is inserted only
at the vertex labeled by $\tau_l$.
The contributions of all the other time orderings may be brought
to this same form by suitably renaming the time variables.

To facilitate the evaluation of the time integral
${\cal I} (\{ k^0_{\alpha} \}, i\nu)$,
consider the sequence of
integrals
\begin{equation}
{\cal I}_n (\tau)
\equiv
\int_{0}^{\tau} d\tau_{n} \, e^{-\sigma_n\tau_n}
\int_{0}^{\tau_n} d\tau_{n-1} \, e^{-\sigma_{n-1}\tau_{n-1}}
\cdots
\int_{0}^{\tau_2} d\tau_1 \, e^{-\sigma_1\tau_1}
\;,
\label{eq:defineIn}
\end{equation}
with the initial condition ${\cal I}_0 (\tau) \equiv 1$.
The original time integral is
${\cal I} (\{ k^0_{\alpha} \}, i\nu)
=
{\cal I}_n (\beta)$.
The functions ${\cal I}_n (\tau)$ satisfy
\begin{equation}
{d\over d\tau} {\cal I}_n (\tau)
= e^{-\sigma_n\tau} {\cal I}_{n-1}(\tau)
\;.
\label{eq:differentialrel}
\end{equation}
This differential relation implies
a simple algebraic recursion relation
for the Laplace transforms of the ${\cal I}_n (\tau)$,
\begin{eqnarray}
{\cal L}_n (\lambda) & \displaystyle \equiv & \displaystyle
\int_0^{\infty} d\tau \,
e^{-\lambda\tau} \,
{\cal I}_n (\tau)
=
\int_0^{\infty} d\tau \,
\left(
{-1\over\lambda}{d\over d\tau} e^{-\lambda\tau}
\right) \,
{\cal I}_n (\tau)
\nonumber\\
& \displaystyle = & \displaystyle
{1\over \lambda}
\vphantom{\Biggl( \Biggr)}
\int_0^{\infty} d\tau \,
e^{-\lambda\tau} \,
{d\over d\tau} {\cal I}_n (\tau)
=
{1\over \lambda}
\int_0^{\infty} d\tau \,
e^{-\lambda\tau - \sigma_n\tau} \,
{\cal I}_{n-1} (\tau)
\nonumber\\
& \displaystyle = & \displaystyle
{1\over\lambda}
\vphantom{\Biggl( \Biggr)}
{\cal L}_{n-1} (\lambda+\sigma_n)
\;.
\label{eq:recursion}
\end{eqnarray}
Iterating this equation produces
\begin{eqnarray}
{\cal L}_n (\lambda)
& \displaystyle = & \displaystyle
{1\over \lambda}
{1\over (\lambda+\sigma_n)}
{1\over (\lambda+\sigma_n + \sigma_{n-1})}
\cdots
{1\over (\lambda+\sigma_n +\cdots +\sigma_1)}
\nonumber\\
& \displaystyle = & \displaystyle
{1\over \lambda}
\prod_{j=1}^n
{1\over (\lambda + \Omega_j^n)}
\;,
\label{eq:iteration}
\end{eqnarray}
where we have defined
\begin{equation}
\Omega_j^n \equiv \sum_{s=j}^n \sigma_s
\;.
\label{eq:originalomegajn}
\end{equation}
The original integral is recovered by performing
an inverse Laplace transform,
\begin{equation}
{\cal I}_n (\tau) =
\int_{\gamma - i\infty}^{\gamma + i\infty}
{d\lambda \over 2\pi i} \,
e^{\lambda\tau} \,
{\cal L}_n (\lambda)
\;,
\label{eq:inverselaplace}
\end{equation}
where $\gamma > \Omega_j^n$ for all $j \leq n$.
Closing the contour anti-clockwise at infinity
and picking up the residues of all the poles yields
\begin{eqnarray}
{\cal I} (\{ k^0_{\alpha} \}, i\nu)
=
{\cal I}_n (\beta)
& \displaystyle = & \displaystyle
\prod_{j=1}^n {1\over (\Omega_j^n)}
\; + \;
      \sum_{k=1}^{n}
          e^{-\beta\Omega_k^n}
                {1 \over (-\Omega_k^n)}
                     \prod^{n}_{
                              \scriptstyle{j=1}
                                  \atop
                              \scriptstyle{j \neq k}
                               }
                                {1\over (\Omega_j^n - \Omega_k^n)}
\nonumber\\
& \displaystyle = & \displaystyle
\prod_{j=1}^n {1\over (\Omega_j^n)}
\; + \;
      \sum_{k=1}^{n}
          e^{-\beta\Omega_k^n}
              \prod^{k-1}_{
                           \scriptstyle{j=1}
                              \atop
                           \scriptstyle{\vphantom{j \neq k}}
                          }
                           {1\over (\Omega_j^{k-1})}
              \prod^{n+1}_{j'=k+1}
                   { 1\over (-\Omega_k^{j'-1}) }
\nonumber\\
& \displaystyle = & \displaystyle
\sum_{k=1}^{n+1}
    e^{-\beta\Omega_k^n}
        \prod_{j=1}^{k-1} {1\over (\Omega_j^{k-1})}
             \prod_{j'=k+1}^{n+1} {1\over (-\Omega_k^{j'-1})}
\nonumber\\
& \displaystyle = & \displaystyle
\sum_{k=0}^n
    e^{-\beta \Omega_{k+1}^n}
       \left(
             \prod_{j=1}^k (\Omega_{j}^k) \,
                    \cdot \,
                  \prod_{j'=k+1}^n (-\Omega_{k+1}^{j'})
       \right)^{-1}
\;,
\\
\; & \vphantom{where} & \;
\nonumber
\label{eq:premain}
\end{eqnarray}
where
\begin{equation}
\Omega_j^k \equiv
\Omega_j^n - \Omega_{k+1}^n
\;
\end{equation}
and in the last two lines
a product that has no elements is to be
interpreted as $1$ and $e^{-\beta\Omega_{n+1}^n} \equiv 1$.

The time integral ${\cal I}$    
is always well-defined and finite.
However individual terms in the
expression (\ref{eq:premain}) become singular when
any of the factors in the denominator vanish.
It will be convenient to regulate
the expression by adding an infinitesimal imaginary
part $i\epsilon_s$
to each frequency sum $\sigma_s$ so that
none of the denominator factors
vanish for any
values of momenta and frequency.
Any set of infinitesimal imaginary parts
may be used as long as these conditions are met.
Different choices produce
equivalent results for
${\cal I}$,
but create differing
pole structure in individual terms.
Since the rules derived in this section involve each individual term,
a different choice of infinitesimal imaginary parts can lead to a
different set of rules.
The most convenient choice will be one which naturally leads to the
standard Cutkosky rules in the zero temperature limit.

The following choice of the infinitesimal parts
will prove to be convenient.
Define
\begin{equation}
{\cal I} (\{ k^0_{\alpha} \}, i\nu)
=
\lim_{\epsilon \to 0}
\,
\sum_{k=0}^n
       \left(
             \prod_{j=1}^k (\Omega_{j}^k - i\epsilon_j^k) \,
                    \cdot \,
        \prod_{j'=k+1}^n (-\Omega_{k+1}^{j'} + i\epsilon_{k+1}^{j'})
       \right)^{-1}
    \exp \left\{
       \beta (-\Omega_{k+1}^n + i\epsilon_{k+1}^n)
         \right\}
\;,
\label{eq:mainresult}
\end{equation}
where
\begin{eqnarray}
\; &
\epsilon_p^n
\equiv \displaystyle
\left\{
   \begin{array}{ll}
    (n - p + 1) \epsilon \; ; & \quad \mbox{for $l{+}1 \leq p \leq n$,} \\
    (n + p - l) \epsilon \; ; & \quad \mbox{for $1 \leq p \leq l$}
   \end{array}
\right.
& \;
\\
\noalign{\hbox{and}}
\; &
\displaystyle
\quad
\epsilon_j^k
\equiv
\epsilon_j^n - \epsilon_{k+1}^n
\;.
& \;
\label{eq:redefinedepsilon}
\end{eqnarray}
Here, the index $l$ corresponds to the vertex $\tau_l$
where the external frequency is inserted.

For further simplification of Eq.~(\ref{eq:mainresult}),
we need to examine the structure of the
factors in ${\cal I}$ more closely.
In view of the definition Eq.~(\ref{eq:sigmal}),
the factor $\Omega_j^k = \sum_{s=j}^k \sigma_s$
contains the sum of
frequencies of all the lines connected to any of the times
$\{ \tau_j,\tau_{j+1},\cdots,\tau_k \}$ with signs indicating whether the
line comes from an earlier time $(+\omega_{\rm in}^s)$
or a later time $(-\omega_{\rm out}^s)$.
Among these frequencies,
those corresponding to the lines connecting the
times $\{ \tau_j,\tau_{j+1},\cdots,\tau_k \}$ to themselves
cancel in the sum for $\Omega_j^k$.
Hence,
\begin{equation}
{\rm Re}\,
\Omega_j^k
=
\sum_{
\alpha \in \Lambda_{\rm in}^{j,k}
}
k^0_{\alpha}
-
\sum_{
\alpha' \in \Lambda_{\rm out}^{j,k}
}
k^0_{\alpha'}
\;,
\label{eq:omegajk}
\end{equation}
where, as illustrated in Figure~\ref{fig:lambda},
$\Lambda_{\rm in}^{j,k}$ indicates the set of lines connecting
the times $\{ \tau_j,\tau_{j+1},\cdots,\tau_{k} \}$ with the earlier times
$\{ \tau_0,\tau_1,\tau_2,\cdots,\tau_{j-1} \}$,
and $\Lambda_{\rm out}^{j,k}$
indicates a set of lines connecting the times
$\{ \tau_j,\tau_{j+1},\cdots,\tau_k \}$ with
the later times
$\{ \tau_{k+1}, \tau_{k+2},\cdots,\tau_n \}$.
\begin{figure}
\begin{picture}(130, 100)(10,20)
\put (110,110) {time}
\put (145,180) {$\tau_n$}
\put (145,130) {$\tau_k$}
\put (147,120) {$\cdot$}
\put (147,110) {$\cdot$}
\put (147,100) {$\cdot$}
\put (145,90) {$\tau_j$}
\put (145,40) {$\tau_0$}
\put (245,10) {$\Lambda_{\rm in}^{j,k}$}
\put (245,205) {$\Lambda_{\rm out}^{j,k}$}
\end{picture}
\leavevmode\epsfbox{lambda.ps}
\caption{%
\advance\baselineskip by -8pt
Illustration of the sets of lines
$\Lambda_{\rm in}^{j,k}$ and $\Lambda_{\rm out}^{j,k}$.
The shaded box contains the vertices labeled with
$\{ \tau_j,\cdots,\tau_k \}$ and the lines connected to themselves.
The set $\Lambda_{\rm in}^{j,k}$
contains the lines going into the box while
$\Lambda_{\rm out}^{j,k}$ contains those coming out of the box.
}
\label{fig:lambda}
\end{figure}
In particular,
\begin{equation}
{\rm Re} \,
\Omega_{k}^n
=
\sum_{
\alpha \in \Lambda_{\rm in}^{k,n}
}
k^0_{\alpha}
\;,
\label{eq:omegasupern}
\end{equation}
so that the exponential in Eq.~(\ref{eq:mainresult}) has the
form
\begin{equation}
\exp
\Bigl\{
\beta(-\Omega_{k}^n + i\epsilon_k^n)
\Bigr\}
=
\exp
\Bigl\{
i\beta \epsilon^n_{k}
-\sum_{\alpha \in \Lambda_{\rm in}^{k,n}}
\beta k^0_{\alpha}
\Bigr\}
\;.
\end{equation}
In the expression  (\ref{eq:schematicint})
for the result of the complete graph,
the time integral ${\cal I}$ is multiplied by
frequency integrals containing
Bose-Einstein factors
and the spectral densities of free propagators.
Since the Bose-Einstein factors
satisfy
$e^{-\beta\omega} \,(1 {+} n(\omega)) =n(\omega)$,
and the spectral densities are odd functions of the
frequency, the identity
\begin{eqnarray}
\int {d\omega \over 2\pi}\,
\Bigl( 1 + n(\omega) \Bigr)\,
\rho_0 (\omega)\,
e^{-\beta\omega}\, f(\omega)
& \displaystyle = & \displaystyle
\int {d\omega \over 2\pi}\,
n(\omega)\, \rho_0 (\omega)\, f(\omega)
\nonumber\\
& \displaystyle = & \displaystyle
\int {d\omega \over 2\pi}\,
\Bigl( 1 + n(\omega) \Bigr)\,
\rho_0 (\omega)\,
f(-\omega)
\;
\label{eq:absorbexp}
\end{eqnarray}
holds for any function $f(\omega)$.
This identity may be used to absorb
the factors of
$e^{-\beta {\rm Re}\, \Omega_{k+1}^n}$
appearing in the result (\ref{eq:mainresult}) for the time integral
${\cal I}$.
When inserted into the expression for the complete graph, the
factor of
$e^{-\beta {\rm Re} \, \Omega_{k+1}^n}$
in each term of ${\cal I}$
may be omitted
if, at the same time, the sign
of all frequencies in the set $\Lambda_{\rm in}^{k+1,n}$ are changed
in the corresponding denominator.\footnote{If the external operators
involve time derivatives, the corresponding factors in the numerator
must also be changed.}

To see the effect of this sign change on the remaining factors of
$\Omega_{j}^{k}$ in the denominator,
recall that ${\rm Re}\, \Omega_{j}^{k}$
is the sum of
the incoming frequencies of the lines
$\Lambda_{\rm in}^{j, k}$
entering the block  of the times
$\{ \tau_{k} , \cdots,  \tau_{j} \}$
minus
the sum of outgoing frequencies
of the the lines $\Lambda_{\rm out}^{j, k}$
leaving the block.
To be a member of the set
$ \Lambda_{\rm out}^{j, k} $, a line must cross the interval just
above the latest time in the time block, $\tau_{k}$.
All lines that cross this interval are also members of
$\Lambda_{\rm in}^{k+1, n}$
since
$\Lambda_{\rm in}^{k+1, n}$
consists of all lines that cross
the same interval.
Thus when the exponential factor
$e^{-\beta {\rm Re}\, \Omega_{k+1}^n}$
is absorbed, all outgoing
frequencies in ${\rm Re}\, \Omega_{j}^{k}$ change sign.
On the other hand, none of
the lines corresponding to the
incoming frequencies for the set of times
$\{ \tau_{k} , \cdots,  \tau_{j} \}$
are members of $\Lambda_{\rm in}^{k+1, n}$ because
these incoming
lines all terminate at times earlier than $\tau_{k+1}$.
Thus the incoming frequencies are not affected by the absorption of
the exponential factor.
Hence, after this change of variables,
${\rm Re}\, \Omega_j^k$
is replaced by the frequency sum
\begin{equation}
\Lambda_j^k
\equiv
\sum_{
     \alpha \in \Lambda_{\rm in}^{j,k}
      }
       k^0_{\alpha}
+
\sum_{
        \alpha' \in \Lambda_{\rm out}^{j,k}
       }
        k^0_{\alpha'}
\;.
\end{equation}
Note that
$\Lambda_j^k$
is the sum of frequencies of all
lines that connect any of the times
$\{ \tau_k,~\tau_{k-1},~\cdots,~\tau_j \}$
to times outside of this set.
A similar result holds for the factors involving
${\rm Re}\, \Omega_{k+1}^{j'}$.
In this case, it
is the incoming frequencies that change sign and
${\rm Re}\, \Omega_{k+1}^{j'}$ is replaced by
\begin{equation}
-\Lambda_{k+1}^{j'}
\equiv
     -\sum_{
           \alpha \in \Lambda_{\rm in}^{k+1,j'}
           }
            k^0_{\alpha}
     -\sum_{
           \alpha' \in \Lambda_{\rm out}^{k+1,j'}
           }
            k^0_{\alpha'}
\;,
\label{eq:Lambda*}
\end{equation}
which is minus the sum of frequencies of all
lines that connect any of the times
$\{ \tau_{j'},~\tau_{j'-1},~\cdots,~\tau_{k+1} \}$
to times outside of this set.
%
%

Consequently, the expression (\ref{eq:mainresult})
for the result of the time integration
may be replaced by the equivalent result
\begin{equation}
{\cal I} (\{ k^0_{\alpha} \}, i\nu)
=
\sum_{k=0}^n\,
S_k (\{ k^0_{\alpha} \}, i\nu)
\,
e^{i\beta\epsilon_{k+1}^n}
\;,
\label{eq:reducedmain}
\end{equation}
where,
for $n \geq k \geq l$,
\addtocounter{equation}{1}
$$
S_k (\{ k^0_{\alpha} \}, i\nu)
\equiv
    \prod^{l}_{j=1}
       \Bigl(\Lambda_j^k -i\nu - i\epsilon_j^k \Bigr)^{-1}
    \prod^{k}_{j'=l+1}
       \Bigl(\Lambda_{j'}^{k} - i\epsilon_{j'}^{k} \Bigr)^{-1}
    \prod^{n}_{j''=k+1}
       \Bigl(\Lambda_{k+1}^{j''} + i\epsilon_{k+1}^{j''} \Bigr)^{-1}
\;,
\eqno{\theequation a}
\label{eq:summanda}
$$
and
$$
S_k (\{ k^0_{\alpha} \}, i\nu)
\equiv
    \prod^{k}_{j=1}
       \Bigl(\Lambda_j^k - i\epsilon_j^k \Bigr)^{-1}
         \prod^{l-1}_{j'=k+1}
            \Bigl(\Lambda_{k+1}^{j'} + i\epsilon_{k+1}^{j'} \Bigr)^{-1}
   \prod^{n}_{j''=l}
    \Bigl(\Lambda_{k+1}^{j''} +i\nu + i\epsilon_{k+1}^{j''} \Bigr)^{-1}
\;,
\eqno{\theequation b}
\label{eq:summandb}
$$
for $l{-}1 \geq k \geq 0$.

The exponential factor
$e^{ i \beta\epsilon^n_{k+1} }$
in Eq.~(\ref{eq:reducedmain})
cannot be omitted
if two or more lines share the same energy.
In that case,
the spectral densities of the propagators may cause some of the
factors in the denominator of $S_k$
to vanish (except for the $i\epsilon$'s) for all
values of the spatial
loop momenta.
Then terms of order $\epsilon$ and higher
in the expansion of the
exponential can result in finite contributions, and hence
must be kept.
For the purpose of deriving simple diagrammatic rules, this is
an inconvenience.
To avoid this,
note that
if the external 4-momentum is non-zero,
only self-energy insertions can cause multiple lines to have identical
4-momentum.
If all self-energy insertions are resummed, thereby replacing bare
propagators by fully dressed one-particle propagators, then
no self-energy insertions remain and
the factor
$e^{i\beta\epsilon_{k+1}^n}$ may be set to one.
The full propagator is given by
\begin{equation}
\tilde{G}^E (|{\bf k}|, i\nu)
\equiv
{
                1
              \over
\nu^2 + E_k^2 + \Sigma(|{\bf k}|, i\nu)
}
\;,
\label{eq:fullpropagator}
\end{equation}
where $ \Sigma( |{\bf k}|, i\nu ) $ is the full one-particle
self-energy in the imaginary-time formulation.
The spectral density of this propagator is given by
\begin{eqnarray}
i \rho ( |{\bf k}|, \omega )
& \equiv & \displaystyle
\tilde{G}^E ( |{\bf k}|, \omega {+} i\epsilon)
-
\tilde{G}^E ( |{\bf k}|, \omega {-} i\epsilon)
\nonumber\\
\; & = & \displaystyle
\left[
E_k^2
- (\omega {+} i\epsilon)^2
+ \Sigma (|{\bf k}|, \omega {+} i\epsilon)
\right]^{-1}
-
\left[
E_k^2
- (\omega {-} i\epsilon)^2
+ \Sigma (|{\bf k}|, \omega {-} i\epsilon)
\right]^{-1}
\;,
\nonumber\\
\;
\label{eq:sdofgl}
\end{eqnarray}
and the corresponding ``mixed'' propagator is
\begin{equation}
\tilde{G}^E ( |{\bf k}|, \tau )
=
\int {d\omega \over 2\pi} \, \Bigl( 1 + n(\omega) \Bigr) \,
\rho ( |{\bf k}|, \omega ) \, e^{-\omega |\tau|}
\;.
\label{eq:mixedfullpropagator}
\end{equation}
\begin{figure}
\begin{picture}(130, 100)(10,20)
\put (145,120) {$\tau_l$}
\put (170,120) {$i\nu$}
\put (180,215) {$\tau_n$}
\put (260,210) {$( \Lambda_n^n - i\epsilon )^{-1}$}
\put (262,191) {$\cdot$}
\put (262,172) {$\cdot$}
\put (262,153) {$\cdot$}
\put (260,135) {$( \Lambda_{l+1}^n - i (n{-}l)\epsilon )^{-1}$}
\put (260,110) {$( \Lambda_l^n - i\nu - in\epsilon )^{-1}$}
\put (262,92.5) {$\cdot$}
\put (262,75) {$\cdot$}
\put (262,57.5) {$\cdot$}
\put (260,40) {$( \Lambda_1^n - i\nu - i(n{+}1{-}l)\epsilon )^{-1}$}
\put (180,30) {$\tau_0$}
\put (205,30) {$i\nu$}
\end{picture}
\leavevmode\epsfbox{predisc1.ps}
\caption{%
\advance\baselineskip by -8pt
A schematic illustration of $S_n$.
The denominators from each interval is shown at the right.
The upper shaded box contains times $\tau_{l+1}$ to $\tau_n$ and lines
connecting them. The lower shaded box contains times $\tau_0$ to
$\tau_{l-1}$ and lines connecting them. At the times $\tau_0$ and
$\tau_l$, two external operators are inserted.
The $\Lambda_s^n$ denotes the sum of all frequencies of the
lines crossing the interval between $\tau_{s-1}$ and $\tau_s$.
}
\label{fig:predisc1}
\end{figure}
{}From now on, unless stated explicitly otherwise, the fully dressed
one-particle propagator will be used and
the term ``skeleton expansion'' will denote the
expansion in which the propagators are fully dressed but the
vertices are not.
Of course, for a practical calculation, some
approximation of the full propagator must be made.
Because of infrared divergences,
performing a consistent expansion in the powers of the coupling
constant may not be completely
trivial \cite{Kapusta}.
This will be discussed further in Section \ref{sec:applications}.

We return now to the original time integral
(\ref{eq:reducedschematic}).
Recall that the integral
${\cal I}$ was one of the $n!$ different
time orderings in the complete integral
$\cal J$.
As shown
in Eq.~(\ref{eq:reducedmain}), each different time ordering generates
a sum of $n{+}1$ individual terms.
Hence, there are a total of $(n{+}1)!$
terms contributing to $\cal J$.
These correspond precisely
to the $(n{+}1)!$ different time orderings of
the $n{+}1$ vertices in the original graph (including the one vertex
whose time was set to zero).
In particular, each term in the result
(\ref{eq:reducedmain}) is naturally associated with a cyclic
permutation of the $n{+}1$ times,
$\tau_{n}~\geq~\cdots~\geq~\tau_1~\geq~0$.
This may be seen as follows.
Consider the real part of the factors
$\Lambda_j^k$ and $\Lambda_{k+1}^{j'}$
appearing in the denominator of $S_k$.
As noted earlier,
$\Lambda_j^k$
is the sum of frequencies of all
lines that connect any of the times
$\{ \tau_k,~\tau_{k-1},~\cdots,~\tau_j \}$
to times outside of this set.
Equivalently,
when the time ordering starts with the sequence
$\tau_k \geq \tau_{k-1} \geq \cdots \geq \tau_j \geq \cdots$,
$\Lambda_j^k$
is the sum of frequencies of all lines
crossing the interval preceding $\tau_j$,
since for this ordering, any line
connected to any of the times
$\{ \tau_k,~\tau_{k-1},~\cdots,~\tau_j \}$ from outside this set
must cross the
interval just below the earliest time in the set, $\tau_j$.
Similarly, $\Lambda_{k+1}^{j''}$
is the sum of frequencies of all
lines that connect any of the times
$\{ \tau_{j''},~\cdots,~\tau_{k+2},~\tau_{k+1} \}$
to times outside of this set.
Equivalently,
it can be interpreted as the sum of frequencies of all lines
crossing the
interval just above $\tau_{j''}$ when the time ordering
ends with
$\cdots \geq \tau_{j''} \geq \cdots \geq \tau_{k+2} \geq \tau_{k+1}$.
In this case,
a line
connected to any of the times
$\{ \tau_{j''},~\cdots,~\tau_{k+2},~\tau_{k+1} \}$ from outside this set
must cross the
interval just above the largest time in the set, $\tau_{j''}$.

Since $j$ ranges from $1$ to $k$ and $j''$ ranges from $k{+}1$ to $n$,
these completely determine the time ordering,
$\tau_k \geq \tau_{k-1} \geq \cdots \geq \tau_1 \geq 0
\geq \tau_n \geq \tau_{n-1} \geq \cdots \geq \tau_{k+1}$.
Hence, each denominator
factor of $\Lambda_j^k$ and $\Lambda_{k+1}^{j''}$ can
be interpreted as the sum of frequencies of all lines crossing each
interval in this time ordering.
The external frequency contributions have a similar interpretation.
When $\tau_l > \tau_0$,
the frequency denominators of $S_k$ corresponding to
intervals between $\tau_l$ and
$\tau_0 \equiv 0$
contain $-i\nu$.
\begin{figure} 
\begin{picture}(130, 100)(10,20)
\put (165,160) {$\tau_l$}
\put (190,160) {$i\nu$}
\put (195,80) {$\tau_n$}
\put (305,215) {$( \Lambda_k^k - i\epsilon )^{-1}$}
\put (307,205) {$\cdot$}
\put (307,194) {$\cdot$}
\put (307,183) {$\cdot$}
\put (305,172) {$( \Lambda_{l+1}^k - i(k{-}l)\epsilon )^{-1}$}
\put (305,145) {$( \Lambda^{k}_l - i\nu - ik\epsilon )^{-1}$}
\put (307,135) {$\cdot$}
\put (307,125) {$\cdot$}
\put (307,115) {$\cdot$}
\put (305,105) {$( \Lambda_1^k - i\nu - i(k{+}1{-}l)\epsilon )^{-1}$}
\put (305,70) {$( \Lambda_{k+1}^n + i(n{-}k)\epsilon )^{-1}$}
\put (307,60) {$\cdot$}
\put (307,50) {$\cdot$}
\put (307,40) {$\cdot$}
\put (305,30) {$( \Lambda_{k+1}^{k+1} + i\epsilon )^{-1}$}
\put (195,100) {$\tau_0$}
\put (220,100) {$i\nu$}
\end{picture}
\leavevmode\epsfbox{predisc2.ps}
\caption{%
\advance\baselineskip by -8pt
A schematic illustration of $S_k$ with $n > k \geq l$.
This diagram is
cyclically related to the diagram in Figure~\ref{fig:predisc1}.
The denominators from each interval are shown at the right.
The upper shaded box contains times $\tau_{l+1}$ to $\tau_{k}$
and lines connected to them.
The middle shaded box contains times $\tau_{0}$ to $\tau_{l-1}$
and lines connected to them.
The lower shaded box contains times $\tau_{k+1}$ to $\tau_{n}$
and lines connected to them.
}
\label{fig:predisc2}
\end{figure}
Since external frequency $i\nu$ flows out of the vertex labeled with
$\tau_l$ and flows into the vertex labeled with
$\tau_0$,
the contribution
$-i\nu$ is always minus the sum of external frequencies flowing out of
the graph from above the interval.
A similar interpretation is also possible when
$\tau_0 > \tau_l$. In this case,
the denominator factors of $S_k$ corresponding to
the intervals between $\tau_0$ and $\tau_l$ contain $+i\nu$.
Since the ordering of $\tau_l$ and $\tau_0$ is reversed compared to
the previous case, again $+i\nu$ equals
the sum of
external frequencies flowing out of the vertices above the interval.
Finally,
this interpretation also holds for intervals
which are not between the two external vertices.
The part of the diagram above both $\tau_0$ and $\tau_l$ has no
external vertex above it,
thus the net external frequency flow is zero;
the part of the diagram below both $\tau_0$ and $\tau_l$ has two
external vertices above it,
and the net external frequency flow is
again zero.
\begin{figure} 
\begin{picture}(130, 100)(10,20)
\put (155,60) {$\tau_l$}
\put (180,60) {$i\nu$}
\put (172,124) {$\tau_n$}
\put (182,148) {$\tau_0$}
\put (207,148) {$i\nu$}
\put (275,200) {$( \Lambda_k^k + i\epsilon )^{-1}$}
\put (277,189) {$\cdot$}
\put (277,178) {$\cdot$}
\put (277,167) {$\cdot$}
\put (275,156) {$( \Lambda_{1}^k + ik\epsilon )^{-1}$}
\put (275,125) {$( \Lambda_{k+1}^n + i\nu
                   + i(n{+}k{+}1{-}l)\epsilon )^{-1}$}
\put (277,115) {$\cdot$}
\put (277,105) {$\cdot$}
\put (277,95) {$\cdot$}
\put (275,85) {$( \Lambda_{k+1}^l + i\nu + i(k{+}1)\epsilon )^{-1}$}
\put (275,55) {$( \Lambda_{k+1}^{l-1} - i(l{-}k{-}1)\epsilon )^{-1}$}
\put (277,45) {$\cdot$}
\put (277,35) {$\cdot$}
\put (277,25) {$\cdot$}
\put (275,15) {$( \Lambda_{k+1}^{k+1} - i\epsilon )^{-1}$}
\end{picture}
\leavevmode\epsfbox{predisc3.ps}
\caption{%
\advance\baselineskip by -8pt
A schematic illustration of
$S_k$ with $l > k \geq 0$.
This is also
cyclically related to the diagram in Figure~\ref{fig:predisc1}.
The denominators from each interval are shown explicitly at the right.
The upper shaded box contains times $\tau_{0}$ to $\tau_{k}$
and lines connected to them.
The middle shaded box contains times $\tau_{l+1}$ to $\tau_{n}$
and lines connected to them.
The lower shaded box contains times $\tau_{k+1}$ to $\tau_{l-1}$
and lines connected to them.
}
\label{fig:predisc3}
\end{figure}

In this way, each contribution
$S_k$ is associated with the cyclic
permutation of the $n{+}1$ times,
$\tau_{n}~\geq~\cdots~\geq~\tau_1~\geq~0$.
Figure~\ref{fig:predisc1} schematically illustrates $S_n$.
Figure~\ref{fig:predisc2}
shows a term $S_k$ with $n > k \geq l$,
while Figure~\ref{fig:predisc3} shows $S_k$ with
$l > k \geq 0$.

The assignment of infinitesimal
$i\epsilon$'s in the denominator of the result
(\ref{eq:mainresult})
may seem complicated. But only the
following two properties are necessary.
First,
the contributions of
intervals above the vertex $\tau_l$
when $\tau_l > \tau_0$, as well as
the contributions of
intervals below $\tau_l$
when $\tau_0 > \tau_l$,
all have
negative imaginary infinitesimals, while
contributions of
intervals below the vertex $\tau_0$
when $\tau_l > \tau_0$,
and the contributions of
intervals above $\tau_0$
when $\tau_0 > \tau_l$,
all have
positive imaginary infinitesimals.
Note that the contributions from these intervals do not contain
the external frequency $i\nu$.
Since the factors without $i\nu$ are not affected by
taking the discontinuity across the real line in the complex
$i\nu$ plane,
only the signs of imaginary
infinitesimals in the contributions from these intervals matter,
but not their relative amplitudes.
Second, between $\tau_l$ and $\tau_0$,
starting from $\tau_l$, the infinitesimal imaginary part
increases
in each successive interval.
These two conditions will be needed to simplify the final result
and relate it to the standard Cutkosky rules.

In summary, in this section the following has been shown.
When evaluating a Feynman diagram in the imaginary-time
formulation, the intermediate
frequency summations can be performed
before any other calculations.
Each term of
the result may be naturally associated with an
old-fashioned time-ordered perturbation theory diagram.
The contribution of each time-ordered diagram is given by
the product of frequency denominators from each time intervals
multiplied by a
statistical factor and spectral density for each line.
Hence a given Feynman diagram
$\Gamma$
generates a contribution of the form
\begin{eqnarray}
{C}^{AB}_{\Gamma} ({\bf q}, i\nu)
& \equiv & \displaystyle
{ (-\lambda)^{n-1}
\over
{\cal S}_{\Gamma} }
\int
\prod_{L \in \Gamma} {d^3 {\bf k}_L \over (2\pi)^3} \,
\prod_{\alpha \in \Gamma }
{dk^0_{\alpha} \over 2\pi} \,
\Bigl( 1 + n(k^0_{\alpha}) \Bigr)
\rho(|{\bf k}_{\alpha}| , k^0_{\alpha})
\nonumber\\
\; & \; & \displaystyle
\qquad\qquad\qquad
{}\times
f^{AB}( \{ {\bf k}_L \}, {\bf q} )
\left(
\sum_{{\Gamma_{\sigma}} \subset \Gamma} \;
S_{{\Gamma_{\sigma}}} (\{ k^0_{\alpha} \}, i\nu)
\right)
\;,
\label{eq:summary2}
\end{eqnarray}
where
${\Gamma_{\sigma}}$
labels the different time-ordered diagrams and
$S_{{\Gamma_{\sigma}}}$
is the product of frequency denominators
\begin{equation}
S_{{\Gamma_{\sigma}}}
\equiv
\prod_{
{
\hbox{\scriptsize{\rm intervals}}
\atop
\hbox{\scriptsize{\it j}}
}
}
\Bigl(
\Lambda_j -  i\nu_j - i\epsilon_j
\Bigr)^{-1}
\;.
\label{eq:redefinedSk}
\end{equation}
Here,
$\Lambda_j$ is
the sum of frequencies of all lines crossing the given
interval,
and $i\nu_j$ is
the net external frequency flowing out of the diagram
above the given interval.
The vertices in a given time-ordered diagram can always be
chronologically relabeled so that the largest time is $\tau_n$ and the
smallest $\tau_0$.
Then
\begin{eqnarray}
\; & \displaystyle
\Lambda_j
=
\sum_{ \alpha \in \Lambda_{\rm in}^{j,n} }
k^0_{\alpha}
& \;
\\
\noalign{\hbox{and}}
\; &
\displaystyle
i\nu_j
=
i\nu
\Bigl(
\theta(l-j + 1/2)
-
\theta(m-j + 1/2)
\Bigr)
\;,
& \;
\end{eqnarray}
where
$\tau_l$ labels the vertex where external frequency flows out
(corresponding to the insertion of the operator $\hat{A}$)
and $\tau_m$ labels the vertex where the external frequency flows in
(corresponding to the insertion of the operator $\hat{B}$).
As before,
the subscript $\alpha$ labels the different lines,
$L$ labels the different loops,
and
${\cal S}_{\Gamma}$ denotes the overall symmetry factor
associated with the diagram.
The factor
$f^{AB}( \{ {\bf k}_L \}, {\bf q} )$
represents the action of the external operators.
Assigning
the infinitesimal imaginary part $i\epsilon_j$ is a little
complicated.
For diagrams with $\tau_l > \tau_m$,
(or where operator $\hat{A}$ acts after $\hat{B}$),
\begin{equation}
\epsilon_j
\equiv
\left\{
   \begin{array}{ll}
    \epsilon \; ; & \quad \mbox{for $l{+}1 \leq j \leq n$,} \\
    (j - m) \epsilon \; ; & \quad
                              \mbox{for $m{+}1 \leq j \leq l$,}\\
    -\epsilon \; ; & \quad \mbox{for $1 \leq j \leq m$,}
   \end{array}
\right.
\end{equation}
while for diagrams with $\tau_m > \tau_l$ the assignment are reversed,
\begin{equation}
\epsilon_j
\equiv
\left\{
   \begin{array}{ll}
    -\epsilon \; ; & \quad \mbox{for $m{+}1 \leq j \leq n$,} \\
    (m - j + 1) \epsilon \; ; & \quad \mbox{for $l{+}1 \leq j \leq m$}\\
    \epsilon \; ; & \quad \mbox{for $1 \leq j \leq l$ \,\,.}
   \end{array}
\right.
\end{equation}
Between $\tau_l$ and $\tau_m$,
starting from $\tau_l$, the magnitude of $\epsilon_j$
deceases in each successive interval.
For intervals before, or after, the action of both operators, only the
sign of the imaginary part $\epsilon_j$ will matter.

\narrowtext
\section{evaluating the discontinuity}
\label{sec:discontinuity}

To find
the spectral density, one must take
the discontinuity
of the analytically continued two-point function
across the real axis
in the complex frequency plane
\begin{equation}
i\rho_{AB} ({\bf q}, \omega)
\equiv
\tilde{G}_{AB}^E ({\bf q}, \omega+i\epsilon)
-
\tilde{G}_{AB}^E ({\bf q}, \omega-i\epsilon)
\;.
\end{equation}
Once
the imaginary-time integrations are done,
performing
the analytic continuation in the external frequency is trivial.
Furthermore,
it is possible to take the discontinuity before
any spatial loop momenta integrals are evaluated.

The only source of a discontinuity comes from
the product of energy denominators
denoted by $S_{{\Gamma_{\sigma}}}$ in the previous result
(\ref{eq:summary2}).
To evaluate
the discontinuity, it is convenient first to make a partial fraction
expansion of $S_{{\Gamma_{\sigma}}}$.
This simply means
repeatedly using the identity
\begin{equation}
{1 \over XY}
=
\left(
{1 \over Y}
-
{1 \over X}
\right)
{1 \over X - Y}
\;
\end{equation}
to write $S_{{\Gamma_{\sigma}}}$ as a sum of simple poles in the
external frequency.
The standard identity
\begin{equation}
{\rm Disc}\, {1\over x}
=
{1\over x+i\epsilon}
-
{1\over x-i\epsilon}
=
-2\pi i \delta (x)
\;
\end{equation}
may then be applied.
If
the vertex $\tau_l$ where the external frequency flows out is
above the vertex $\tau_m$ where the external frequency flows in,
one obtains
\begin{eqnarray}
{\rm Disc}\, S_{{\Gamma_{\sigma}}} (\{ k^0_{\alpha} \}, \omega)
& \displaystyle \equiv & \displaystyle
S_{{\Gamma_{\sigma}}} (\{ k^0_{\alpha} \}, \omega + i\epsilon)
-
S_{{\Gamma_{\sigma}}} (\{ k^0_{\alpha} \}, \omega - i\epsilon)
\nonumber\\
& \displaystyle = & \displaystyle
\sum_{c=m+1}^l
\prod_{j=l+1}^n
   \Bigl( \Lambda_j
         - i \epsilon \Bigr)^{-1}
\prod_{p=c+1}^l
   \Bigl( \Lambda_p - \omega
         - i (\epsilon_p - \epsilon_c) \Bigr)^{-1}
\nonumber\\
\; & \; & \displaystyle
\; {}\times
(2\pi i) \delta(\Lambda_c - \omega)
\prod_{q=m+1}^{c-1}
   \Bigl( \Lambda_q - \omega
         + i (\epsilon_c - \epsilon_q) \Bigr)^{-1}
\prod_{j'=1}^m
   \Bigl(\Lambda_{j'}
         + i\epsilon \Bigr)^{-1}
\;.
\nonumber\\
\eqnum{\theequation a}
\label{eq:discofSk1}
\end{eqnarray}
Alternatively, if
the vertex $\tau_l$ where the external frequency flows out is
below the vertex $\tau_m$ where the external frequency flows in,
the discontinuity is given by
\begin{eqnarray}
{\rm Disc}\, S_{{\Gamma_{\sigma}}} (\{ k^0_{\alpha} \}, \omega)
& \displaystyle \equiv & \displaystyle
\sum_{c=l+1}^m
\prod_{j=m+1}^n
   \Bigl( \Lambda_j
         + i \epsilon \Bigr)^{-1}
\prod_{p=c+1}^m
   \Bigl( \Lambda_p + \omega
         + i (\epsilon_c - \epsilon_p) \Bigr)^{-1}
\nonumber\\
\; & \; & \displaystyle
\; {}\times
(-2\pi i) \delta(\Lambda_c + \omega)
\prod_{q=l+1}^{c-1}
   \Bigl( \Lambda_q+ \omega
         - i (\epsilon_q - \epsilon_c) \Bigr)^{-1}
\prod_{j'=1}^l
   \Bigl(\Lambda_{j'}
         - i\epsilon \Bigr)^{-1}
\;.
\nonumber\\
\eqnum{\theequation b}
\label{eq:discofSk2}
\end{eqnarray}
\addtocounter{equation}{1}
\begin{figure} 
\begin{picture}(130, 100)(10,20)
\put (170,185) {$\tau_l$}
\put (200,185) {$\omega$}
\put (200,80) {$\tau_m$}
\put (235,80) {$\omega$}
\put (317,209) {($\Lambda_j - i\epsilon)^{-1}$}
\put (317,157) {($\Lambda_p - \omega - i\epsilon)^{-1}$}
\put (317,135) {$2\pi\delta(\Lambda_c - \omega)$}
\put (317,104) {($\Lambda_q - \omega + i\epsilon)^{-1}$}
\put (317,47) {($\Lambda_{j'} + i\epsilon)^{-1}$}
\end{picture}
\leavevmode\epsfbox{disc2.ps}
\caption{%
\advance\baselineskip by -8pt
A schematic illustration of a time-ordered cut diagram
${\Gamma\llap{$-$}}_{\sigma}$
corresponding to the ${\rm Disc}\, S_{{\Gamma_{\sigma}}}$
in Eq.~(\ref{eq:discofSk1})
where the external frequency flows out above the cut.
The vertices are chronologically ordered so that the largest time
is $\tau_n$ and the smallest $\tau_0$,
and
$n \geq j \geq l{+}1$,
$l \geq p \geq c{+}1$,
$c{-}1 \geq q \geq m{+}1$, and
$m \geq j' \geq 1$.
The dashed line indicates the cut and
the contribution of all the intervals are shown at the right.
}
\label{fig:disc2}
\end{figure}
The sum (with index $c$) runs over the different poles in external
frequency of $S_{\Gamma_{\sigma}}$, or equivalently, enumerates each
interval between the two external vertices.
Typical diagrams corresponding to these results are illustrated in
Figure~\ref{fig:disc2} and \ref{fig:disc3}.
After taking the discontinuity,
only the signs of the infinitesimal imaginary part matters;
the relative magnitudes of
$\epsilon_j$'s are no longer relevant.
The combinations of $\epsilon_j$'s
appearing in the above
expression are all positive.
Thus every infinitesimal in the above expression may be
replaced by the same $i\epsilon$.

Each term of the discontinuity
is naturally associated
with a ``cut'' time-ordered diagram.
This is a time-ordered
diagram with a line,
called the ``cut'',
drawn across one time interval
which must lie
between the two external vertices.
The result (\theequation)
is written in such way that the
contributions of the later intervals in a cut diagram
appear earlier in the expression.
\begin{figure} 
\begin{picture}(130, 100)(10,20)
\put (165,63) {$\tau_l$}
\put (195,63) {$\omega$}
\put (195,188) {$\tau_m$}
\put (230,188) {$\omega$}
\put (312,204) {($\Lambda_j + i\epsilon)^{-1}$}
\put (312,147) {($\Lambda_p+ \omega + i\epsilon)^{-1}$}
\put (312,118) {$-2\pi\delta(\Lambda_c + \omega)$}
\put (312,89) {($\Lambda_q + \omega - i\epsilon)^{-1}$}
\put (312,37) {($\Lambda_{j'} - i\epsilon)^{-1}$}
\end{picture}
\leavevmode\epsfbox{disc3.ps}
\caption{%
\advance\baselineskip by -8pt
A schematic illustration of a time-ordered cut diagram
${\Gamma\llap{$-$}}_{\sigma}$
corresponding to the ${\rm Disc}\, S_{{\Gamma_{\sigma}}}$ in
Eq.~(\ref{eq:discofSk2})
where the external frequency flows out below the cut.
The vertices are chronologically ordered so that the largest time
is $\tau_n$ and the smallest $\tau_0$,
and
$n \geq j \geq m{+}1$,
$m \geq p \geq c{+}1$,
$c{-}1 \geq q \geq l{+}1$, and
$l \geq j' \geq 1$.
The dashed line indicates the cut and
the contribution of all the intervals are shown at the right.
}
\label{fig:disc3}
\end{figure}
The complete contribution of each cut diagram may obviously be
factored to the product form,
\begin{equation}
{\rm Disc}\, S_{{\Gamma_{\sigma}}} ( \{k^0_{\alpha}\}, \omega )
=
i s
\sum_{{{\Gamma\llap{${\scriptstyle -}$}}_{\sigma}}
\subset {\Gamma_{\sigma}}} \,
2\pi
\delta(
\sum_{
{
\hbox{\scriptsize{\rm cut lines}}
\atop
\hbox{\scriptsize{\it c}}
}
}
k^0_c - s\omega )\,
S_{{{\Gamma\llap{${\scriptstyle -}$}}_{\sigma}}}^{+}
      ( \{k^0_{\alpha}\}, \omega ) \,
S_{{{\Gamma\llap{${\scriptstyle -}$}}_{\sigma}}}^{-}
      ( \{k^0_{\alpha}\}, \omega )
\;,
\end{equation}
where the sum runs over all cut time-ordered diagrams, indicated by
${\Gamma\llap{$-$}}_{\sigma}$,
generated by the original time-ordered diagram
$\Gamma_{\sigma}$, and
\begin{equation}
S_{{{\Gamma\llap{${\scriptstyle -}$}}_{\sigma}}}^{+}
( \{k^0_{\alpha}\}, \omega )
\equiv
\prod_{p=n-n_{+}+1}^n
   \Bigl( \Lambda_p - \omega_p
         - i s\epsilon \Bigr)^{-1}
\;,
\end{equation}
denotes the contribution from the upper part of the cut diagram,
and
\begin{equation}
S_{{{\Gamma\llap{${\scriptstyle -}$}}_{\sigma}}}^{-}
( \{k^0_{\alpha}\}, \omega )
\equiv
\prod_{p=1}^{n_{-}}
   \Bigl( \Lambda_p - \omega_p
         + i s\epsilon \Bigr)^{-1}
\;
\end{equation}
denotes the contribution from the lower part.
Here, $n_{+}$ is the number of
interaction vertices above the cut,
$n_{-}$ is the
number of
interaction vertices below the cut,
and
$s$ is $+1$ if the external frequency enters
lower part of the diagram,
$-1$ otherwise.
$\omega_p$ is the net (real continuous) external frequency flowing
out of the vertices above the $p$-th interval
and equals
$s\omega$ for intervals between the external vertices, and
$0$ otherwise.
The contribution of each interval is summarized by the rules:
\begin{enumerate}
\item Each uncut time
interval between two vertices contributes a factor
which is
the inverse of the sum of frequencies of all lines crossing the
interval, minus the total external frequency flowing out of the
vertices above the interval,
plus or minus $i\epsilon$.
If the interval is on the side where the external
frequency flows out,
use $-i\epsilon$, if on the other side, use
$+i\epsilon$.

\item The cut interval contributes
$2\pi$ times a delta function whose argument is the sum of the
frequencies of all lines
crossing the interval minus the external frequency
flowing out of the vertex above
the cut.
The sign of the delta function is
$(+)$ if the cut is above the vertex where the external frequency
flows in
and $(-)$ if below.
\end{enumerate}

The fact
that each side of the cut interval contributes the energy denominators
with same sign of $i\epsilon$ is important,
and will enable
these cut time-ordered
diagrams to be resummed
into the simpler
cut Feynman
diagrams considered in the next section.
Note that
if the contribution of an interval $j$ in a cut time-ordered
diagram is given by
$( \Lambda_j {-} \omega {\pm} i\epsilon )^{-1}$,
then the contribution of the corresponding interval in the
time-reversed diagram is
$( \Lambda_j {+} \omega {\pm} i\epsilon )^{-1}$.
Thus,
the contributions from a time-ordered diagram and
its time-reversal are
related to each other by an
overall sign change combined with the flipping the external frequency,
$\omega \to -\omega$.

Combining these results, the discontinuity of
a complete Feynman diagram
$\Gamma$ can be expressed as a sum over all possible cuts,
\begin{equation}
{\rm Disc}\, {C}^{AB}_{\Gamma} (q)
\equiv
\sum_{{\Gamma\llap{${\scriptstyle -}$}}
\subset \Gamma}
\left(
D^{AB}_{{\Gamma\llap{${\scriptstyle -}$}}} ({\bf q}, q^0)
-
D^{AB}_{{\Gamma\llap{${\scriptstyle -}$}}} ({\bf q}, -q^0)
\right)
\;,
\end{equation}
where
the sum runs over all topologically distinct cut Feynman diagrams,
denoted by
${\Gamma\llap{$-$}}$,
and
$D^{AB}_{{\Gamma\llap{${\scriptstyle -}$}}} ({\bf q}, q^0)$
is
the sum of contributions from all topologically equivalent
cut time-ordered
diagrams with a common cut and with the external frequency flowing out
above the cut,
\begin{eqnarray}
D^{AB}_{{\Gamma\llap{${\scriptstyle -}$}}} (q)
& = & \displaystyle
{ (-\lambda)^{n-1}
\over
{\cal S}_{\Gamma} }
\int
\prod_{L \in \Gamma} {d^3 {\bf k}_L \over (2\pi)^3} \,
\prod_{\alpha \in \Gamma}
{dk^0_{\alpha} \over 2\pi} \,
\Bigl( 1 + n(k^0_{\alpha}) \Bigr)
\rho(|{\bf k}_{\alpha}| , k^0_{\alpha})
f^{AB}( \{ {\bf k}_L \}, {\bf q} )
\nonumber\\
\; & \; & \displaystyle
\qquad\qquad
{}\times
\left(
\sum_{{{\Gamma\llap{${\scriptstyle -}$}}_{\sigma}}
\subset {\Gamma\llap{${\scriptstyle -}$}}}
2\pi i \delta(
\sum_{
{
\hbox{\scriptsize{\rm cut lines}}
\atop
\hbox{\scriptsize{\it c}}
}
}
k^0_c - q^0
)\,
S_{{{\Gamma\llap{${\scriptstyle -}$}}_{\sigma}}}^{+}
(\{k^0_{\alpha}\}, q^0 ) \,
S_{{{\Gamma\llap{${\scriptstyle -}$}}_{\sigma}}}^{-}
(\{k^0_{\alpha}\}, q^0 )
\right)
\;.
\label{eq:summary3}
\end{eqnarray}
Obtaining a more compact form of this result
is the goal of the next section.

\narrowtext
\section{simple cutting rules}
\label{sec:rules}

For diagrams with $n$ vertices,
the number of cut time-ordered diagrams is more than $n!$ times
greater than the
number of original Feynman diagrams.
This is obviously inconvenient.
Fortunately, one may combine different cut time-ordered diagrams
so that each side of the cut yields ordinary Feynman diagrams.

Consider a set of of topologically equivalent cut time-ordered
diagrams with $n{-}1$ interaction vertices.
This means diagrams with the same connectivity and the same cut,
but with any relative ordering of the vertices on each side of the
cut.
Assume, for convenience,
that the external operator $\hat{A}$ emits the external momentum $q$
above the cut.
Let ${\Gamma\llap{$-$}}$ denote
this set of cut time-ordered diagrams
(or equivalently, a single cut Feynman diagram).
As shown in the last section,
the sum of contributions from the graphs in the set ${\Gamma\llap{$-$}}$
is given by
$D^{AB}_{{\Gamma\llap{${\scriptstyle -}$}}} (q)$
(\ref{eq:summary3}).
For each choice of the cut interval, the contribution
$D_{{\Gamma\llap{${\scriptstyle -}$}}}^{AB}(q)$
sums up only half of the cut time-ordered diagrams.
The remaining cut time-ordered diagrams are those where
the external momentum flows out below the cut.
These diagrams, as explained in the previous section,
sum to
$- D_{{\Gamma\llap{${\scriptstyle -}$}}}^{AB} ({\bf q}, -q^0)$.
Below, it will also be shown that
\begin{equation}
D_{{\Gamma\llap{${\scriptstyle -}$}}}^{AB}
({\bf q}, -q^0)
=
e^{-q^0\beta} \, D_{{\Gamma\llap{${\scriptstyle -}$}}}^{AB}
({\bf q}, q^0)
\;.
\label{eq:Drelation}
\end{equation}
Consequently,
the spectral density $\rho_{AB}(q)$ has the diagrammatic expansion
\begin{equation}
i\rho_{AB}(q)
=
\sum_{{\Gamma\llap{${\scriptstyle -}$}}} \,
(1 - e^{-q^0\beta})\,
D_{{\Gamma\llap{${\scriptstyle -}$}}}^{AB}(q)
\;,
\label{eq:sdintermsofsingleD}
\end{equation}
where the sum runs over all topologically distinct
cut diagrams ${\Gamma\llap{$-$}}$.

We wish to
rewrite the expression for
$D_{{\Gamma\llap{${\scriptstyle -}$}}}^{AB} (q)$
in Eq.~(\ref{eq:summary3})
as a product of two contributions representing each half of the cut
diagram, with each contribution produced by
the standard Feynman rules.
This will generate a finite temperature generalization of
the zero temperature Cutkosky rules.
To proceed,
note that
the cut separates the vertices of the diagram into two disjoint sets,
the upper part and lower part.
Thus, the summation over time-orderings, with the cut fixed,
can be separated into
two independent summations over permutations of vertices on each side
of the cut.
Therefore the total contribution from
a cut diagram
${\Gamma\llap{$-$}}$ can be
written as
\begin{eqnarray}
{D}_{{\Gamma\llap{${\scriptstyle -}$}}}^{AB} (q)
& = & \displaystyle
{i \over {\cal S}_{\Gamma}}
\int {d^4{q'} \over (2\pi)^4} \,
\prod_{
{
\hbox{\scriptsize{\rm cut lines}}
\atop
\hbox{\scriptsize{\it c}}
}
}
\left(
{d^4 k_{c} \over (2\pi)^4} \,
\Bigl( 1 + n(k^0_{c}) \Bigr)
\rho( k_{c} )
\right)
\nonumber\\
\; & \; & \displaystyle
\quad \quad \quad
{}
\times
D^{A}_{\Gamma^{+}}
( \{ k_{c} \}, q ) \,
D^{B}_{\Gamma^{-}}
( \{ k_{c} \}, {q'} ) \,
\;,
\end{eqnarray}
where
$D^{A}_{\Gamma^{+}}( \{ k_{c} \}, q )$
and
$D^{B}_{\Gamma^{-}}( \{ k_{c} \}, {q'} ) \,$
denote the results of resummation in the upper $(\Gamma^{+})$ and
the lower $(\Gamma^{-})$ parts of the cut diagram, respectively.
Here and henceforth
the subscript $c$ labels
the different cut lines.
An extra integration over ${q'}$ is introduced because each
factor
$D^{{\cal O}^{\pm}}_{\Gamma^{\pm}}$
(with ${\cal O}^{+} \equiv \hat{A}$
and ${\cal O}^{-} \equiv \hat{B}$)
will contain
an overall
4-momentum conserving delta function.
Hence their product
will produce
an overall factor of
$(2\pi)^4 \delta^4 (q - q')$
which is removed by the integration over $q'$.

Explicitly, the contributions from each half of the diagram are given
by
\begin{eqnarray}
D^{{\cal O}^{\pm}}_{\Gamma^{\pm}}
( \{ k_{c} \}, q )
& \equiv & \displaystyle
(-\lambda)^{n_{\pm}}
(2\pi)^3 \delta^3 ( \sum_{c} {\bf k}_{c} - {\bf q} )
\int \prod_{L \in \Gamma^{\pm}}
{d^3 {\bf k}_{L} \over (2\pi)^3} \,
\prod_{ \alpha \in \Gamma^{\pm} }
{dk^0_{\alpha} \over 2\pi} \,
\Bigl( 1 + n(k^0_{\alpha}) \Bigr)
\rho(|{\bf k}_{\alpha}| , k^0_{\alpha})
\nonumber\\
\; & \; & \displaystyle
\qquad\qquad\qquad\qquad\qquad\qquad
\quad {}\times
f^{{\cal O}^{\pm}}( \{ {\bf k}_L \}, {\bf q} ) \,
{\cal J}^{\pm} ( \{ k^0_{\alpha} \}, \{ k^0_{c} \}, q^0 )
\;.
\end{eqnarray}
The factor
$f^{{\cal O}^{\pm}}( \{ {\bf k}_L \}, {\bf q} )$ represents the action
of a single external operator.
The factors
${\cal J}^{\pm} ( \{ k^0_{\alpha} \}, \{ k^0_c \}, q^0 )$
denote the sum
of the products of the frequency denominators
over the permutations of the vertices on each side of
the cut,
\begin{equation}
{\cal J}^{\pm} ( \{ k^0_{\alpha} \}, \{ k^0_c \}, q^0 )
\equiv
2\pi \delta(
        \sum_{c}
               k^0_c - q^0)
\sum_{ {\Gamma\llap{${\scriptstyle -}$}_{\sigma}} \subset
\Gamma^{\pm} }
S_{{{\Gamma\llap{${\scriptstyle -}$}}_{\sigma}}}^{\pm}
    ( \{ k^0_{\alpha} \}, \{ k^0_c \}, q^0 )
\;,
\label{eq:cutsum}
\end{equation}
where as before
\begin{eqnarray}
\; &
\displaystyle
S_{{{\Gamma\llap{${\scriptstyle -}$}}_{\sigma}}}^{+}
    ( \{ k^0_{\alpha} \}, \{ k^0_c \}, \omega )
\equiv
\prod_{j=n-n_{+}+1}^n
\Bigl( \Lambda_j - \omega_j
                  - i\epsilon \Bigr)^{-1}
\;,
& \;
\label{eq:S+}
\\
\noalign{\hbox{and}}
\; &
\displaystyle
S_{{{\Gamma\llap{${\scriptstyle -}$}}_{\sigma}}}^{-}
    ( \{ k^0_{\alpha} \}, \{ k^0_c \}, \omega )
\equiv
\prod_{j=1}^{n_{-}}
\Bigl( \Lambda_j - \omega_j
                  + i\epsilon \Bigr)^{-1}
\;.
& \;
\label{eq:S-}
\end{eqnarray}
Here, it should be understood that the summation is over the
permutations of vertices on each side of the cut.
Recall that
the times are labeled in chronological
order so that the largest time is always labeled as
$\tau_n$ for each time ordering.
Thus, in any given time-ordering,
$\Lambda_j {-} \omega_j$ can be replaced by
$\Omega_j^n = \sum_{s=j}^n \sigma_s$
({\it c.f.} Eq.~(\ref{eq:originalomegajn})).

For the upper part of the diagram,
the product of frequency denominators
in $S_{{{\Gamma\llap{${\scriptstyle -}$}}_{\sigma}}}^{+}$
(\ref{eq:S+}),
together with the overall
frequency conserving delta function, has exactly the same
form as the contribution of a zero temperature
time-ordered diagram corresponding to the upper part of the cut
diagram.
Consequently,
$S_{{{\Gamma\llap{${\scriptstyle -}$}}_{\sigma}}}^{+}
( \{ k^0_{\alpha} \}, \{ k^0_c \}, q^0 )$
times a delta function
equals the ordered real-time integral,
\begin{eqnarray}
\lefteqn{
2\pi \delta (\Omega_{n-n_{+}}^n) \,
S_{{{\Gamma\llap{${\scriptstyle -}$}}_{\sigma}}}^{+}
( \{ k^0_{\alpha} \}, \{ k^0_c \}, q^0 )
=}
\nonumber\\
& &
\displaystyle
\quad\qquad\qquad
(i)^{n_{+}}
\int_{-\infty}^{\infty} dt_{n-n_{+}} \,
e^{-i\sigma_{n-n_{+}} t_{n-n_{+}}}
\int_{t_{n-n_{+}}}^{\infty} dt_{n-n_{+}+1} \,
e^{-i\sigma_{n-n_{+}+1} t_{n-n_{+}+1}}
\nonumber\\
& &
\displaystyle
\quad\qquad\qquad
\quad\cdots
\int_{t_{n-1}}^{\infty} dt_n \,
e^{-i\sigma_n t_n}
\;,
\end{eqnarray}
which is simply the
$t_n \geq t_{n-1} \geq \cdots \geq t_{n-n_{+}}$
piece of the unordered real-time integral
\begin{equation}
{\cal J}^{+}( q^0, \{ k^0_{\alpha} \}, \{ k^0_c \} )
=
(i)^{n_{+}}
\int_{-\infty}^{\infty}
\prod_{p=n-n_{+}}^n dt_p \,
e^{i q^0 t_l}
\,
\prod_{
{
\hbox{\scriptsize{\rm cut lines}}
\atop
\hbox{\scriptsize{\it c}}
}
}
\!\! e^{-ik^0_c t_d}
\,
\prod_{
{
\hbox{\scriptsize{\rm uncut lines}}
\atop
\hbox{$\scriptstyle \alpha$}
}
}
\!\! e^{-i(k^0_{\alpha} - i\epsilon) |t_a - t_b|}
\;
\label{eq:J+}
\end{equation}
corresponding to the same half of the cut diagram.
As previously,
$t_l$ denotes the vertex where the external frequency
$q^0$ flows out
and all the times
$\{ t_a, t_b, t_d, t_l \}$
are within the set
$\{t_{n-n_{+}}, \cdots, t_n \}$.
The frequency of the cut lines are regarded as flowing into the upper
part of the diagram; hence each cut line contributes
a factor of $e^{-ik^0_c t_d}$, while the external line contributes a
factor of $e^{i q^0 t_l}$.
The equality of the sum (\ref{eq:cutsum}) and the unordered time
integral (\ref{eq:J+}) can be easily established
since the unordered integral
${\cal J}^{+}(q^0, \{k^0_{\alpha}\}, \{k^0_c\})$
precisely generates
the sum of contributions from all orderings of vertices above the cut.
For future use note that
changing the integration variables
$\{ t_p \}$ to $\{ -t_p \}$ shows that
\begin{equation}
{\cal J}^{+}(q^0, \{k^0_{\alpha}\}, \{k^0_c\})
=
{\cal J}^{+}(-q^0, \{k^0_{\alpha}\}, \{-k^0_c\})
\;.
\end{equation}
Thus, for both
${\cal J}^{+}( \{ k^0_{\alpha} \}, \{ k^0_c \}, q^0 )$
and
$D_{\Gamma^{+}}^{{\cal O}^{+}}( \{ k_{c} \}, q^0 )$,
changing the sign of the external frequency is equivalent to
changing the sign of the frequencies of all cut lines.
This fact will be needed to prove the relation (\ref{eq:Drelation}).

The complete contribution
$D_{\Gamma^{+}}^{{\cal O}^{+}} ( \{ k_{c} \}, q )$
of the upper half of the cut diagram
can thus be produced by applying standard Feynman rules
using the simple real-time propagator
\begin{equation}
G ( |{\bf k}|, |t-t'| )
=
\int {d\omega \over 2\pi} \,
\Bigl( 1 + n(\omega) \Bigr)
\rho (|{\bf k}|, \omega)
e^{-i\omega |t-t'|}
\;,
\end{equation}
or equivalently, using its Fourier transform
\begin{eqnarray}
\tilde{G} (k)
& \equiv & \displaystyle
\int dt \,
e^{ik^0 t}
G ( |{\bf k}|, |t| )
\nonumber\\
\; & = & \displaystyle
\int {d\omega \over 2\pi i} \,
\Bigl( 1 + n(\omega) \Bigr)
\left(
{
          1
        \over
k^0 + \omega - i\epsilon
}
-
{
          1
        \over
k^0 - \omega + i\epsilon
}
\right)
\rho (|{\bf k}|, \omega)
\nonumber\\
\; & = & \displaystyle
\int {d\omega \over 2\pi} \,
\Bigl( 1 + n(\omega) \Bigr)
\left(
{
       2i\omega
        \over
(k^0)^2 - (\omega - i\epsilon)^2
}
\right)
\rho( |{\bf k}|, \omega)
\;,
\label{eq:realtimeprop}
\end{eqnarray}
for each uncut line, and regarding each cut line as
an external line with definite frequency and momentum.
Explicitly, this result is
\begin{equation}
D_{\Gamma^{+}}^{{\cal O}^{+}}
( \{ k_{c} \}, q )
=
(-i\lambda)^{n_{+}}
(2\pi)^4 \delta^4 ( \sum_{c} k_{c} - q )
\int
\prod_{L \in \Gamma^{+}}
{ d^4 k_L \over (2\pi)^4 } \,
f^{{\cal O}^{+}} ( \{ {\bf k}_L \}, {\bf q} )
\prod_{\alpha \in \Gamma^{+}}
\tilde{G} ( k_{\alpha} )
\;.
\end{equation}

The same procedure may be applied to the lower part of the cut
diagram. Due to the difference in sign of the infinitesimal imaginary
parts, the result involves complex conjugated propagators.
Noting that
$\Omega_j^n {=} -\Omega_0^{j-1}$,
the contribution of a particular time-ordering
$\tau_{n_{-}} \geq \tau_{n_{-}-1} \geq \cdots \geq \tau_1 \geq 0$
can be expressed by the real-time integral
\begin{eqnarray}
\lefteqn{
2\pi \delta (\Omega_0^{n_{-}})
S_{{{\Gamma\llap{${\scriptstyle -}$}}_{\sigma}}}^{-}
    ( \{ k^0_{\alpha} \}, \{ k^0_c \}, q^0 ) =}
\nonumber\\
& &
\displaystyle
\quad\qquad\qquad
(-i)^{n_{-}}
\int_{\infty}^{\infty} dt_{n_{-}} \,
e^{i\sigma_{n_{-}} t_{n_{-}}}
\int^{t_{n_{-}}}_{-\infty} dt_{n_{-}-1} \,
e^{i\sigma_{n_{-}-1} t_{n_{-}-1}}
\cdots
\int^{t_{1}}_{-\infty} dt_0 \,
e^{i\sigma_0 t_0}
\;.
\label{eq:I-}
\end{eqnarray}
As before,
summing over all time-orderings of the vertices below the
cut produces
\begin{equation}
{\cal J}^{-}( {q}^0,\{ k^0_{\alpha} \},\{ k^0_c \} )
=
(-i)^{n_{-}}
\int_{-\infty}^{\infty}
\prod_{q=0}^{n_{-}} dt_q \,
e^{-i {q}^0 t_{m}}
\prod_{
{
\hbox{\scriptsize{\rm cut lines}}
\atop
\hbox{\scriptsize{\it c}}
}
}
\!\! e^{ik^0_c t_{g}}
\,
\prod_{
{
\hbox{\scriptsize{\rm uncut lines}}
\atop
\hbox{$\scriptstyle \alpha$}
}
}
\!\! e^{i (k^0_{\alpha} + i\epsilon) |t_{e} - t_{f}|}
\;,
\label{eq:Lambda-}
\end{equation}
where in going from (\ref{eq:I-}) to (\ref{eq:Lambda-})
the integration variables $\{ t_q \}$ are changed to $\{ -t_q \}$.
The time
$t_{m}$ denotes the vertex where the external frequency ${q}^0$
flows in and
the times
$\{ t_{e}, t_{f}, t_{g}, t_{m} \}$
are now within the set
$\{ t_0, \cdots, t_{n_{-}} \}$.
The factors in
${\cal J}^{-}( {q}^0, \{ k^0_{\alpha} \}, \{ k^0_c \} )$
are complex conjugated compared to those in
${\cal J}^{+}( q^0, \{ k^0_{\alpha} \}, \{ k^0_c \} )$.
Thus, the complete contribution from the lower part of the cut diagram
is produced by applying standard Feynman rules, but using complex
conjugated propagators.
Once again, both
${\cal J}^{-}({q}^0, \{k^0_{\alpha}\}, \{k^0_c\})$
and
$D_{\Gamma^{-}}^{{\cal O}^{-}} ( \{ k_{c} \}, q )$
are invariant under simultaneous changes in sign of $q^0$ and the cut
frequencies.
The final contribution of the lower part is
\begin{equation}
D_{\Gamma^{-}}^{{\cal O}^{-}}
( \{ k_{c} \}, q )
=
(i\lambda)^{n_{-}}
(2\pi)^4 \delta^4 ( \sum_{c} k_{c} - q )
\int
\prod_{L \in \Gamma^{-}}
{ d^4 k_L \over (2\pi)^4 } \,
f^{{\cal O}^{-}} ( \{ {\bf k}_L \}, {\bf q} )
\prod_{\alpha' \in \Gamma^{-}}
\tilde{G} ( k_{\alpha'} )^*
\;.
\end{equation}

Combining both halves,
the complete expression for the finite temperature cut diagram,
defined as the sum of all topologically equivalent cut time-ordered
diagrams with the cut corresponding to
$2\pi \delta ( \sum_{c} k^0_c - q^0 )$,
is given by
\begin{eqnarray}
D_{{\Gamma\llap{${\scriptstyle -}$}}}^{AB} (q)
& = & \displaystyle
{
i(-i\lambda)^{n_{+}} (i\lambda)^{n_{-}}
                \over
          {\cal S}_{\Gamma}
}
\int
\prod_{L \in \Gamma}
{d^4 k_{L} \over (2\pi)^4} \,
\prod_{
{
\hbox{\scriptsize{\rm cut lines}}
\atop
\hbox{\scriptsize{\it c}}
}
}
\Bigl( 1 + n(k^0_{c}) \Bigr)
\rho( k_{c} )
\nonumber\\
\; & \; & \displaystyle
\qquad\qquad\qquad\qquad
\quad {}\times
f^{AB} ( \{ {\bf k}_L \}, {\bf q} ) \,
\prod_{\alpha \in \Gamma^{+}}
\tilde{G} (k_{\alpha}) \,
\prod_{\alpha' \in \Gamma^{-}}
\tilde{G} (k_{\alpha'})^*
\;,
\end{eqnarray}
using the fact that
after integrating over the delta functions, the remaining cut momenta
are all independent loop momenta.
This expression for the finite temperature cut diagrams is
almost identical to that of zero temperature cut diagrams
given by the Cutkosky rules.
The cut may be regarded as dividing a diagram into shaded and unshaded
regions.
The unshaded region is above the cut
where the external operator $\hat{A}$ emits the external momentum
$q$.
An uncut line with the 4-momentum $k$ in the unshaded region
represents the propagator $\tilde{G}(k)$.
An uncut line with 4-momentum $p$
in the shaded region represents the complex conjugated propagator
$\tilde{G}(p)^*$.
A cut line with 4-momentum $q$ flowing into the unshaded region
represents the phase space factor $(1 {+} n(q^0)) \rho(q)$.

Once again,
note that only cut time-ordered diagrams
with the external momentum flowing out above the cut enter
the definition of the finite temperature cut diagram.
The sum of the remaining cut time-ordered
diagrams where the external momentum flows out below the cut
are not independent and
can be expressed in terms of the same finite temperature
cut diagrams.
As stated earlier,
the sum of cut time-ordered diagrams where the external
momentum flows out below the cut equals
\begin{eqnarray}
-
D^{AB}_{{\Gamma\llap{${\scriptstyle -}$}}} ( {\bf q}, -q^0 )
& = & \displaystyle
{-i
\over
{\cal S}_{\Gamma}
}
\int {d^4{q'} \over (2\pi)^4} \,
\prod_{
{
\hbox{\scriptsize{\rm cut lines}}
\atop
\hbox{\scriptsize{\it c}}
}
}
\left(
{d^4 k_{c} \over (2\pi)^4} \,
\Bigl( 1 + n(k^0_{c}) \Bigr)
\rho( k_{c} )
\right)
\nonumber\\
\; & \; & \displaystyle \qquad {}\times
D^{A}_{\Gamma^{+}}
( \{ k_{c} \}, ( {\bf q}, -q^0 )) \,
D^{B}_{\Gamma^{-}}
( \{ k_{c} \}, ( {\bf q}',-{q'}^0 )) \,
\;.
\end{eqnarray}
Both of the factors
$D^{{\cal O}^{\pm}}_{\Gamma^{\pm}}$
are invariant under simultaneous sign changes of both
the external frequency and the frequencies of the cut lines.
Thus, changing the integration variables
$\{ k^0_{c} \}$
to
$\{ -k^0_{c} \}$
easily yields
\begin{equation}
D^{AB}_{{\Gamma\llap{${\scriptstyle -}$}}} ( {\bf q}, -q^0 )
=
e^{-q^0\beta} \,
D^{AB}_{{\Gamma\llap{${\scriptstyle -}$}}} ( {\bf q}, q^0 )
\;,
\label{eq:remainder}
\end{equation}
after using the relation
$( 1 {+} n(-k^0) ) \rho( |{\bf k}|, -k^0 )
=
e^{-k^0\beta}( 1 {+} n(k^0) ) \rho( |{\bf k}|, k^0 )$
and overall frequency conservation.

The relations (\ref{eq:remainder}) and (\ref{eq:sdintermsofsingleD})
show that
the finite temperature cut diagrams
$D^{AB}_{{\Gamma\llap{${\scriptstyle -}$}}} (q)$
directly generate
the perturbative expansion of the correlation function
\begin{equation}
i\sigma_{AB}(q)
\equiv
i\int dx \, e^{-iqx}
\langle \hat{A}(x) \hat{B}(0) \rangle
\;
\end{equation}
which is related to the spectral density by the
the standard relation~\cite{Fetter,Brown},
\begin{eqnarray}
\rho_{AB}(q)
=
(1 - e^{-q^0\beta}) \, \sigma_{AB}(q)
\;.
\label{eq:sdintermsofsigma}
\end{eqnarray}
Furthermore,
the definition of the spectral density as a commutator, or
\begin{equation}
\rho_{AB}(q)
=
\sigma_{AB}(q) - \sigma_{BA}(-q)
\;,
\end{equation}
combined with the identity (\ref{eq:sdintermsofsigma}),
implies that
the sum of all cut time-ordered diagrams with the external momentum
$q$ flowing out below the cut yields $-\sigma_{BA}(-q)$.
Thus $\sigma_{BA}(-q)$ can be interpreted as
the sum of all finite temperature cut diagrams where the external
operator $\hat{A}$ emits the momentum $q$ from the {\it shaded}
region whereas $\sigma_{AB}(q)$ is given by cut diagrams where
operator $\hat{A}$ is in the {\it unshaded} region.

Finite temperature cut diagrams differ from
zero temperature cut diagrams in one notable way.
At zero temperature, each side of a cut
diagram must itself be a connected diagram.
But at non-zero temperature,
the cut need only separate the two external vertices.\footnote{
%
%
Thus, for each diagram with a total of $n$ interaction vertices, the
number of finite temperature cut diagrams (neglecting symmetry
factors)
is $2^{n}$.}
Although the complete diagram must be connected, each half of the cut
diagram, considered separately, may have multiple disconnected parts.
A typical such diagram is illustrated in
Figure~\ref{fig:disconnected}.
\begin{figure}%
\begin{picture}(100, 100)(10, 20)
\end{picture}
\leavevmode\epsfbox{disconnected.ps}
\caption{%
\advance\baselineskip by -8pt
A typical finite temperature cut diagram which is absent at zero
temperature.
}
\label{fig:disconnected}
\end{figure}
This change reflects the fact that at non-zero temperature,
a perturbation can lower the internal energy of the original thermal
ensemble.
At the diagrammatic level, this means that a cut line can have
a 4-momentum with either positive or negative energy,
and
so a collection of physical excitations can have
vanishing net 4-momentum.
Consequently, a disconnected piece in half of a cut diagram can produce
a non-zero ``on-shell'' result at finite temperature.

Given the previous results,
the diagrammatic rules for the spectral density can
be cited in two equivalent ways.
One may use
finite temperature cut diagrams to calculate
the correlation function
$\sigma_{AB}(q)$
and then simply multiply it by
$(1 - e^{-q^0\beta})$
to obtain the spectral density.
Or, one may use
finite temperature cut diagrams to calculate both
$\sigma_{AB}(q)$ and $\sigma_{BA}(-q)$ separately and
subtract them to yield the spectral density.
The following diagrammatic
rules summarize the latter approach which directly
reflects the commutator origin of the
spectral density.
Rules for the first method are obtained simply
by restricting the set of diagrams
to those in which the external momentum flows out of the unshaded
region of the cut diagram.
These rules also apply when the external operators involve time
derivatives of the elementary fields.
A sketch of the proof is contained in Appendix~\ref{app:derivatives}.
To calculate the perturbative expansion of a finite temperature
spectral density:

\begin{enumerate}
\item Draw all suitable topologically distinct cut diagrams
(including a choice of shading).
Label each line with a 4-momentum $k$,
assign the external 4-momentum,
and conserve 4-momentum at each vertex.

\item On the unshaded side of the cut,
apply the standard Feynman rules; use the propagator
$
\tilde{G} (k)
$
({\it c.f.} Eq.~(\ref{eq:realtimeprop}))
for each uncut line and
assign a factor of $-i\lambda$ to each interaction vertex.

\item On the shaded side,
apply complex conjugated Feynman rules;
use the propagator
$ \tilde{G} (k)^* $
for each uncut line, and
assign $i\lambda$ to each interaction vertex.

\item Each cut line with momentum $k$ flowing from the shaded
region into the unshaded region contributes a thermal phase space
factor
$$
\tilde{\Delta} (k)
\equiv
\Bigl( 1+n(k^0) \Bigr) \rho (k)
\;.
$$

\item Integrate over all loop momenta, and divide by the symmetry
factor. If the external frequency flows out of the shaded region,
assign an additional factor of $-1$.
\end{enumerate}

As the temperature goes to zero, the phase space factor
$\tilde{\Delta} (k)$
develops
a step function
$\theta(k^0) \left. \rho (k) \right|_{T=0}$
which forces all cut lines to carry positive energy.
Inserting
the spectral density of the bare propagator
reduces $\tilde{\Delta} (k_{c})$ to
$\theta(k^0_{c}) \delta(k^2_{c}+m^2)$
and
immediately reduces
these finite temperature rules to the standard zero
temperature Cutkosky rules.

\narrowtext
\section{applications}
\label{sec:applications}

To illustrate the application of the cutting rules,
consider the spectral density of the correlation function
$\langle {\cal T} ({\hat{\phi}}^{2} ({\bf x}, t)
\hat{\phi}^2 (0) )\rangle$
in a $\lambda\phi^4$ theory.
In particular, we will focus on
the quantity
\begin{equation}
\eta_{\phi^2 \phi^2} \equiv
\lim_{\omega \to 0}
\lim_{{\bf k} \to 0}
\left[\,
{
\rho_{\phi^2 \phi^2} ({\bf k}, \omega)
\over \omega
}
\,\right]
\;.
\end{equation}
This differs from the shear viscosity (\ref{eq:viscosity}),
only by the omission of spatial derivatives in the operator
insertions.
Including these derivatives would make only minor notational changes
in the following discussion.
When the Boltzmann equation is valid,
the transport coefficients can be shown to be inversely proportional
to the scattering cross section.
Finite order perturbative calculations of the transport coefficient
can never produce that result, and hence a partial resummation of
diagrams from all orders will always be necessary.
The example discussed here will show similar behavior.

We will consider the
first few terms in the expansion of $\rho_{\phi^2 \phi^2}$
and analyze the resulting low momentum behavior.
The cutting rules will be used to
evaluate cut diagrams where the external momentum $q$
flows out of the unshaded region.
Multiplying the result with $(1-e^{-q^0\beta})$
will yield
the complete
contribution to the spectral density, as discussed in the
previous section.

The first contribution to
the spectral density comes from the one loop cut
diagram in Figure~\ref{fig:zeroorder}.
\begin{figure}
\begin{picture}(170, 100)(10,20)
\put (260,152) {$-k{+}\omega$}
\put (218,190) {$\omega$}
\put (218,28) {$\omega$}
\put (193,60) {$k$}
\end{picture}
\leavevmode\epsfbox{zeroorder.ps}
\caption{%
\advance\baselineskip by -8pt
The lowest order diagram for the spectral density of
$\langle {\cal T} ({\hat{\phi}}^{2} ({\bf x}, t)
\hat{\phi}^2 (0)) \rangle$
at zero external 3-momentum.
}
\label{fig:zeroorder}
\end{figure}
Following the rules given in the last section,
the cut lines contribute factors of
$( 1 {+} n(k^0))\rho(|{\bf k}|, k^0 )$ and
$( 1 {+} n( -k^0{+}\omega ) ) \rho( |{\bf k}|, -k^0{+}\omega )$
where $\rho(|{\bf k}|, k^0)$ is the full one-particle
spectral density
and the spatial external momentum is set to $0$.
Multiplying these contributions by a factor of
$(1 - e^{-\beta\omega})$ yields
\begin{eqnarray}
\rho_{\phi^2 \phi^2}^{\hbox{\scriptsize{1-loop}}} (0,\omega) & = &
\displaystyle
2(1 - e^{-\beta\omega})
\int {d^4 k \over (2\pi)^4}
\Bigl(1 + n(-k^0 + \omega)\Bigr)
\Bigl(1 + n(k^0)\Bigr)
\nonumber\\
\; & \; & \quad \displaystyle
\qquad\qquad\qquad
{}\times
\rho (|{\bf k}|, -k^0 + \omega)
\rho (|{\bf k}|, k^0)
\;.
\label{eq:fullzeroorder}
\end{eqnarray}
(The leading factor of two is an overall symmetry factor.)
For a free theory, the one-particle spectral density
$\rho(k)$ reduces to a delta function,
\begin{equation}
\rho_{\hbox{\scriptsize{free}}} (|{\bf k}|, k^0)
=
{k^0 \over |k^0|}
2\pi
\delta( k^2 + m^2 )
\;,
\label{eq:freezeroorder}
\end{equation}
and the resulting
integral (\ref{eq:fullzeroorder})
agrees with the standard result~\cite{Weldon}.
The delta function in the free one-particle
spectral density
$ \rho_{\hbox{\scriptsize{free}}} (|{\bf k}|, k^0) $
requires that
$k^0 = \pm E_k$,
and
$ \rho_{\hbox{\scriptsize{free}}} (|{\bf k}|, -k^0 {+} \omega) $
requires
$k^0 = \pm E_k {+} \omega$.
When $|\omega|$ is less than twice the mass $m$,
these two conditions cannot be satisfied simultaneously, and
hence
$\eta_{\phi^2 \phi^2}^{\hbox{\scriptsize{free}}}
= \displaystyle{\lim_{\omega \to 0}}
\rho_{\phi^2 \phi^2}^{\hbox{\scriptsize{free}}} (0, \omega)
/ \omega$
vanishes in a massive free theory.
\footnote{
In the massless limit,
the one-loop spectral density
$\rho_{\phi^2 \phi^2}^{\hbox{\scriptsize{free}}} (0 ,\omega)$
in the free theory is non-zero and
can easily be calculated explicitly.
One finds
\begin{eqnarray*}
\lim_{m \to 0}
\rho_{\phi^2 \phi^2}^{\hbox{\scriptsize{free}}} (0,\omega)
& = &
\displaystyle
2(1 - e^{-\beta\omega})
\int
{d^4 k \over (2\pi)^4}\,
{ 1 \over 4|{\bf k}|^2 }
\Bigl( 1 + n(k^0) \Bigr)
\Bigl( 1 + n(-k^0 {+} \omega) \Bigr)
\nonumber\\
\; & \; & \quad\quad \displaystyle
{}\times
\Bigl(
2\pi\delta(k^0 {-} |{\bf k}|)
-
2\pi\delta(k^0 {+} |{\bf k}|)
\Bigr)
\Bigl(
2\pi\delta(-k^0 {+} \omega {-} |{\bf k}|)
-
2\pi\delta(-k^0 {+} \omega {+} |{\bf k}|)
\Bigr)
\nonumber\\
\; & = & \displaystyle
{1\over \pi}
\coth (\beta\omega / 4)
\;.
\label{eq:zeromass}
\end{eqnarray*}
}

This result changes dramatically when the
spectral density has finite width.
If
\begin{equation}
\Sigma (|{\bf k}|, i\nu) \equiv
\tilde{G}^E (|{\bf k}|, i\nu)^{-1}
-
\nu^2
-
E_k^2
\;
\end{equation}
denotes the full self-energy, then
one-particle spectral density is
\begin{equation}
\rho (k)
=
{
      -2\Sigma^I ( k )
          \over
( k^2 + m^2 + \Sigma^R ( k ) )^2
+
( \Sigma^I ( k ) )^2
}
\;,
\end{equation}
where
$\Sigma (k) \equiv \Sigma^R (k) {+} i\Sigma^I (k)$
when $k^0$
approaches the real line from above in the complex $k^0$
plane.
The spectral density has a peak at $k^0 = E_k$ where the single
particle energy $E_k$ is the solution to
$E_k^2 = {\bf k}^2 {+} m^2 {+} \Sigma^R ({\bf k}, E_k)$.
The width of the peak is given by
$\Gamma_k \equiv
- \Sigma^I ({\bf k}, E_k)/ 2E_k$
and for weak coupling this width is
$O(\lambda^2)$
(since the first graph which contributes an imaginary part is the two
loop graph in Figure~\ref{fig:selfenergy}).
For sufficiently small coupling, the frequency variation in
the self-energy is negligible over the width of the peak in the
spectral density.
Hence the single-particle spectral density may be approximated by
\begin{equation}
\rho (k)
\approx
{
      4k^0 \Gamma_k
          \over
( (k^0)^2 - E_k^2 )^2 + 4( k^0 \Gamma_k )^2
}
\;.
\label{eq:approximatesd}
\end{equation}
With a finite width single particle
spectral density,
the integrand in Eq.~(\ref{eq:fullzeroorder})
is smooth for all $\omega$, and
the limit $\omega \to 0$ and
the momentum integral may be freely interchanged.
Hence
\begin{eqnarray}
\eta_{\phi^2 \phi^2}^{\hbox{\scriptsize{1-loop}}}
& \equiv & \displaystyle
\lim_{\omega \to 0}
{
\rho_{\phi^2 \phi^2}^{\hbox{\scriptsize{1-loop}}} (0,\omega)
             \over
            \omega
}
=
2\beta
\int {d^4 k \over (2\pi)^4}\,
n(k^0)
\Bigl(1 + n(k^0)\Bigr)
\Bigl( \rho (|{\bf k}|, k^0) \Bigr)^2
\nonumber\\
\; & \approx & \displaystyle
2\beta
\int {d^3 {\bf k} \over (2\pi)^3} \,
\int {dk^0 \over 2\pi}
n(k^0)
\Bigl(1 + n(k^0)\Bigr)
\left(
{
      4k^0 \Gamma_k
          \over
( (k^0)^2 - E_k^2 )^2 + 4( k^0 \Gamma_k )^2
}
\right)^2
\nonumber\\
\; & = & \displaystyle
\beta
\int {d^3 {\bf k} \over (2\pi)^3} \,
n(E_k)
\Bigl(1 + n(E_k)\Bigr)
{ 1 \over E_k^2 \Gamma_k }
\Bigl( 1 + O(\Gamma_k / E_k) \Bigr)
\;,
\label{eq:limitzeroorder}
\end{eqnarray}
where the prefactor $\beta$ comes from
$\displaystyle{\lim_{\omega \to 0}} [(1 - e^{-\beta\omega})/\omega]$.
Since the width $\Gamma_k$ is $O(\lambda^2)$ for weak coupling,
this approximation to
$\eta_{\phi^2 \phi^2}^{\hbox{\scriptsize{1-loop}}}$
diverges in the weak coupling limit.
This is in accord with the earlier expectation based on
the Boltzmann equation.

The evaluation of the leading contribution to
$\eta_{\phi^2 \phi^2}^{\hbox{\scriptsize{1-loop}}}$
is now
essentially reduced to the evaluation of the width $\Gamma_k$.
The cutting rules can be used to evaluate the width
since the imaginary part of the self-energy is proportional to the
discontinuity in this case.
The self-energy diagram with the least number of interaction vertices
is given in~Figure~\ref{fig:selfenergy}.
%
%
\begin{figure}
\begin{picture}(160, 100)(10,20)
\put (207,190) {$q$}
\put (207,28) {$q$}
\put (178,60) {$k_1$}
\put (227,60) {$k_2$}
\put (254,60) {$q{-}k_1{-}k_2$}
\end{picture}
\leavevmode\epsfbox{selfenergy.ps}
\caption{%
\advance\baselineskip by -8pt
A two-loop self-energy diagram
in $\lambda \phi^4$ theory.
The slashed lines are the amputated external lines.
}
\label{fig:selfenergy}
\end{figure}
Using the result of the previous section,
the contribution from this
diagram is given by
\begin{eqnarray}
\Sigma^{I}_{\hbox{\scriptsize{2-loop}}} (q)
& = & \displaystyle
-{\lambda^2 \over 12}
(1 - e^{-\beta q^0})
\int {d^4 k_1 \over (2\pi)^4} \,
\int {d^4 k_2 \over (2\pi)^4} \,
\Bigl( 1 + n(k_1^0) \Bigr)
\rho (k_1)
\nonumber\\
\; & \; & \quad {}\times
\Bigl( 1 + n(k_2^0) \Bigr)
\rho (k_2)
\Bigl( 1 + n(q^0 - k_1^0 - k_2^0) \Bigr)
\rho (q - k_1 - k_2)
\;.
\label{eq:twoloop}
\end{eqnarray}
The overall factor $12$ comes from the symmetry factor $6$ and a
factor of $2$ difference between the discontinuity and the imaginary
part.
This two loop diagram is the first single particle irreducible
self-energy diagram in the skeleton expansion.
To evaluate Eq.~(\ref{eq:twoloop}),
some approximation to the full single-particle
self-energy $\Sigma(q)$ must be made.
In this case,
one may easily show that approximating
the full spectral density
by the free single-particle
spectral density is adequate to yield the leading
result.

An approximation scheme using dressed propagators
is equivalent to
adding a self-energy term
${\scriptstyle {1 \over 2}} \phi \Sigma \phi$
to the free Lagrangian and subtracting from the interaction term,
\begin{equation}
-{\cal L} =
{1 \over 2}
\phi ( -\partial_{\tau}^2 - \nabla^2 + m^2 + \Sigma ) \phi
+
\left(
{ {1 \over 4!} } \lambda \phi^4
-
{1 \over 2}
\phi \Sigma \phi
\right)
\;.
\label{eq:modlagrangian}
\end{equation}
By using
$( \nu^2 {+} {\bf k}^2 {+} m^2 {+} \Sigma(i\nu, {\bf k}) )^{-1}$
as the basic propagator and treating
$-{\scriptstyle {1 \over 2}} \phi \Sigma \phi$
as an additional interaction,
a rearranged
perturbation series
can be obtained
for any physical quantity.
If $\Sigma$ is the exact self-energy,
then this expansion is what was previously called a
skeleton expansion.
Since the propagator already contains the full self-energy, all
corrections to the propagator
must add up to zero in the skeleton expansion.
Thus,
the sum of contributions from all one particle irreducible skeleton
self-energy diagrams minus the contribution from the additional
interaction $\Sigma$ vanishes.
This is one of the standard Dyson equations for the $\lambda\phi^4$
theory.

If, in the Lagrangian (\ref{eq:modlagrangian}),
one adds and subtracts only the lowest order,
the sum of contributions from all single particle irreducible
self-energy diagrams minus the contribution of the additional
interaction $\Sigma$
will be non-zero but will be of higher order than the lowest order
self-energy.
In the resulting expansion, the lowest order self-energy acts as an
infrared cut-off and renders finite higher order on-shell single
particle reducible diagrams which would normally have been divergent.
Provided that the correct $O(\lambda^2)$ single-particle width appears
in the rearranged propagators, the resulting diagrammatic series will
generate corrections suppressed by positive powers of
the interaction strength $\lambda$
(perhaps modified by logarithms).
Thus, it is sufficient to compute
the leading behavior of (the imaginary part of)
the two-loop self-energy diagram by replacing the single-particle
spectral densities by free spectral densities in the weak coupling
limit.
The actual calculation is somewhat complicated~\cite{Landsman,Haber}.
Here we report
only the leading
result at high temperature $( T \gg m)$ and vanishing spatial
momentum, $q = (m, 0)$.
One finds~\cite{Parwani},
\begin{equation}
\Sigma^{I}_{\hbox{\scriptsize{2-loop}}} (m,0)
=
-{\rm sgn}(m)
{\lambda^2 T^2 \over 768 \pi}
\Bigl( 1 + O( (m / T) \ln (T / m) ) \Bigr)
\;
\label{eq:highTselfenergy}
\end{equation}
and the static one-particle width is
given by
\begin{equation}
\Gamma_{(m,0)}
=
-\Sigma^{I} (m,0) / 2m
=
\lambda^2 T^2 / 1536m\pi
(1 + O(\lambda))
\;.
\end{equation}
\begin{figure}
\begin{picture}(150, 100)(10,20)
\put (242,152) {$-k_2{+}\omega$}
\put (165,152) {$k_2$}
\put (198,190) {$\omega$}
\put (198,28) {$\omega$}
\put (145,60) {$-k_1{+}\omega$}
\put (235,60) {$k_1$}
\put (215,85) {$k_3$}
\put (190,127) {$p$}
\put (290,115) {$p=-k_1{-}k_2{-}k_3{+}\omega$}
\end{picture}
\leavevmode\epsfbox{secondorder.ps}
\caption{%
\advance\baselineskip by -8pt
A cut ``vertex correction'' diagram
contributing to the spectral density of
$\langle {\cal T} ( {\hat{\phi}}^{2} ({\bf x}, t)
\hat{\phi}^2 (0) ) \rangle$
at zero external 3-momentum.
The dashed line indicates the cut.
The contribution of this diagram together with the one with all
internal lines reversed is denoted by
$\rho_{\phi^2 \phi^2}^{\hbox{\scriptsize v}} (0,\omega)$.
}
\label{fig:secondorder}
\end{figure}
Using this result, the leading behavior of
$\eta_{\phi^2 \phi^2}^{\hbox{\scriptsize{1-loop}}}$
in high temperature limit can be found,
\begin{eqnarray}
\eta_{\phi^2 \phi^2}^{\hbox{\scriptsize{1-loop}}}
& = &
\beta
\int {d^3 {\bf k} \over (2\pi)^3} \,
n(E_k)
\Bigl(1 + n(E_k)\Bigr)
{ 1 \over E_k^2 \Gamma_k^{\vphantom{2}} }
\Bigl( 1 + O(\Gamma_k / E_k) \Bigr)
\nonumber\\
\; & = & \displaystyle
{768 \over \lambda^2 T \pi}
\ln(aT/m)
(1 + O(m/T, \lambda^2))
\;.
\label{eq:highTeta}
\end{eqnarray}
The numerical constant $a$ in the logarithm
depends on
the spatial momentum dependence of the
width
$\Gamma_k$.
(For a constant width, $a=2$ \cite{Haber}.)
The appearance of the logarithm reflects the fact that
when $m{=}0$ this integral is logarithmically divergent.
However, even a massless field develops
a finite thermal mass at finite temperature.
In the case of massless high temperature $\lambda \phi^4$ theory,
the one-loop self-energy provides a thermal mass
$m_T = (\lambda/24)^{1\over 2} T$ \cite{Kapusta,Parwani}.

The one-loop result,
$\eta_{\phi^2 \phi^2}^{\hbox{\scriptsize{1-loop}}}$,
is not the only relevant contribution to
$\eta_{\phi^2 \phi^2}$ at this order.
There are other diagrams with more interaction vertices
that also make contribution of the same order as the one-loop
result.
This is because there are additional infrared divergences at finite
temperature,
and the infrared cut-offs provided by the thermal scattering width of
the self-energy
contribute negative powers of $\lambda$.
For example, consider
the ``eye'' diagram shown in Figure~\ref{fig:secondorder}.
Since this is closely
related to the two-loop
vertex correction, we will denote it by
$\rho_{\phi^2 \phi^2}^{\hbox{\scriptsize v}} (0,\omega)$.
Applying the cutting rules,
the graph in the Figure~\ref{fig:secondorder}
together with the one with all internal lines reversed
generates the contribution
\begin{eqnarray}
\lefteqn{
\rho_{\phi^2 \phi^2}^{\hbox{\scriptsize v}} (0,\omega)
=
}
\nonumber\\
& &
\displaystyle
\qquad\qquad
( 1 - e^{-\beta\omega} )\,
\lambda^2
\int
{d^4 k_1 \over (2\pi)^4}\,
{d^4 k_2 \over (2\pi)^4}\,
{d^4 k_3 \over (2\pi)^4}\,
\int
{d\omega_1\over 2\pi}
{d\omega_2\over 2\pi}
\left( 1 + n(-k_1^0 -k_2^0 -k_3^0 + \omega) \right) \,
\nonumber\\
& &
\displaystyle
\qquad\qquad
{}\times
\left( 1 + n(k_1^0) \right) \,
\Bigl( 1 + n(\omega_1) \Bigr) \,
\left( 1 + n(k_2^0) \right) \,
\Bigl( 1 + n(\omega_2) \Bigr) \,
\left( 1 + n(k_3^0) \right) \,
\nonumber\\
& &
\displaystyle
\qquad\qquad
{}\times
\rho(|{\bf k}_1|, k_1^0) \,
\rho(|{\bf k}_1|, \omega_1) \,
\rho(|{\bf k}_2|, k_2^0) \,
\rho(|{\bf k}_2|, \omega_2) \,
\nonumber\\
& &
\displaystyle
\qquad\qquad
{}\times
\rho(|{\bf k}_3|, k_3^0) \,
\rho( |-{\bf k}_1 - {\bf k}_2 - {\bf k}_3|,
-k_1^0 -k_2^0 -k_3^0 + \omega ) \,
\nonumber\\
& &
\displaystyle
\qquad\qquad
{}\times
\left(
{
       (2i\omega_1)
          \over
[ (-k_1^0 + \omega)^2 - (\omega_1 - i\epsilon)^2 ]
}
{
       (2i\omega_2)
          \over
[ (-k_2^0 + \omega)^2 - (\omega_2 - i\epsilon)^2 ]
}
\right)
\;.
\label{eq:secondorder}
\end{eqnarray}

If free one-particle spectral densities are used,
the kinematical constraints require
$|\omega_1| {=} |k_1^0| {=} E_{k_1}$, and
$|\omega_2| {=} |k_2^0| {=} E_{k_2}$.
Under these constraints, the contribution of the uncut lines
in the lower frequency limit becomes
\begin{equation}
{
               (2i\omega_1)
                   \over
     [ (-k_1^0 + \omega)^2 - (\omega_1 - i\epsilon)^2 ]
}
{
               (2i\omega_2)
                   \over
     [ (-k_2^0 + \omega)^2 - (\omega_2 - i\epsilon)^2 ]
}
=
-
{1 \over \omega^2}
{
       \omega_1 \omega_2
            \over
         k^0_1 k^0_2
}
\;.
\label{eq:underconstraints}
\end{equation}
Combined with the $(1 - e^{-\beta\omega})$ prefactor,
this will cause
$\rho_{\phi^2 \phi^2}^{(2), \hbox{\scriptsize v}} (0,\omega)$
diverge linearly as $\omega \to 0$.

The non-zero width of finite temperature
one-particle spectral densities
will cut-off this divergence.
The
two uncut propagators will be almost on shell.
To estimate the contributions of this diagram,
these propagators may be replaced by their maximum
on-shell values, ${\sim1}/E_k \Gamma_k$.
Thus as the external frequency
$\omega$ tends to $0$, the
diagram's leading contribution behaves as
$\omega\lambda^2 / \Gamma_{k_1}\Gamma_{k_2} \sim \omega / \lambda^2$
effectively replacing the
$\omega^2$ denominator
in Eq.~(\ref{eq:underconstraints}) by the product of inverse life times
$\Gamma_{k_1}\Gamma_{k_2}$.
As before,
the factor of $\omega$ comes from the small $\omega$ limit of the
prefactor $(1 - e^{-\beta\omega})$.
Note that this contribution is the same
order in $\lambda$ as the leading one-loop result.
This demonstrates that
the one-loop calculation is not
sufficient to obtain the correct weak coupling, low frequency
behavior.
We will show below that there are an infinite number of
diagrams contributing at the same order in the zero frequency limit.

The leading behavior of the contribution of a general diagram in the
limit of zero external 4-momentum is not difficult to determine. Since
the source of the inverse powers of the coupling is the infrared
cut-off provided by the thermal scattering width of the self-energy,
the weak coupling behavior is directly
related to the infrared
singularity of the diagram when all the lines are interpreted as free
propagators. At zero temperature, there is a standard argument that
asserts that the appearance of on-shell singularities is due either to
the pinching of an integration contour by coalescing poles or to poles
colliding with the end point of a contour~\cite{Coleman}.
A similar argument may be given for the diagrams considered here.
Suppose we approximate the one-particle spectral density by the
Lorentzian form
\begin{equation}
\rho (k)
\sim
{1 \over E_k}
\left(
{ \Gamma
    \over
(k^0 - E_k)^2 + \Gamma^2
}
-
{ \Gamma
    \over
(k^0 + E_k)^2 + \Gamma^2
}
\right)
\;.
\label{eq:Lorentzian}
\end{equation}
In this approximation,
both
$\tilde{G} (|{\bf k}|, k^0)$
and
$\rho (|{\bf k}|, k^0)$
contain four simple poles
in the complex $k^0$ plane
at
$E_k {\pm} i\Gamma$
and
$-E_k {\pm} i\Gamma$.
Thus the products of propagators and spectral densities
$\tilde{G} (|{\bf k}|, k^0)
\tilde{G} (|{\bf k}|, k^0{+}\omega)$,
$\rho (|{\bf k}|, k^0)
\tilde{G} (|{\bf k}|, k^0{+}\omega)$,
and
$\rho (|{\bf k}|, k^0)
\rho (|{\bf k}|, k^0{+}\omega)$
all have poles
within a circle of radius $\Gamma$ as $\omega \to 0$
which
approach the contour from opposite sides.
Therefore,
after performing
the frequency integrations,
a loop containing one of these products will behave as
${\sim} 1 / \Gamma$ as $\omega \to 0$.
All one has to do to determine the leading behavior
as $\omega \to 0$ is
then to count the number of such loops in skeleton expansion diagrams.

As an example,
consider a generalization of the diagram in
Figure~\ref{fig:secondorder}
to the $(2N{+}1)$-loop planar ``ladder'' diagram illustrated in
Figure~\ref{fig:ladder}.
\begin{figure}%
\begin{picture}(150, 100)(10,20)
\put (260,127) {$N$}
\end{picture}
\leavevmode\epsfbox{ladder.ps}
\caption{%
\advance\baselineskip by -8pt
The planar ladder diagram with $N$ rungs
in $\lambda\phi^4$ theory.
}
\label{fig:ladder}
\end{figure}
When the external 4-momentum vanishes, there are $N{+}1$
pairs of lines with identical 4-momentum in this diagram.
Equivalently, there are $N{+}1$ loops each of which behave like
$1 / \Gamma$.
Thus, all cut diagrams for the $(2N{+}1)$-loop
planar ladder diagram in Figure~\ref{fig:ladder}
can generate
$O(\omega\lambda^{2N} / \Gamma^{N+1})$ contributions
in the low frequency limit.
(As always,
a factor of $\omega$ comes from
the prefactor $(1 - e^{-\beta\omega})$.)
Since the single-particle width
is order $\lambda^2$, all
planar graph of this form
contribute at order
$\omega / \lambda^2$ regardless of the number of loops.
(Similar behavior may be found in other processes~\cite{Kapusta}.)
Other $(2N{+}1)$-loop graphs with
crossed loops
(non-planar graphs)
are necessarily of lower order because
they cannot have as many
pairs of the lines sharing the same 4-momentum.~\footnote{
%
%
Strictly speaking, this applies only when the mass is non-zero. If the
mass is zero, the statistical factors can provide additional sources
of divergent behavior.}
Summing all leading contributions is a non
trivial problem.
We hope to pursue this matter in a future paper.

\acknowledgments

Constant encouragement and guidance of
L. G. Yaffe and L. S. Brown are greatly appreciated.

\appendix{$C\!PT$ invariance and the spectral density}
\label{app:cpt}

Consider the consequence of the $C\!PT$ invariance on the
spectral density of the two-point function of any pair of operators
$\hat{A}$ and $\hat{B}$.
Assume that $\hat{A}$ and $\hat{B}$ have definite behavior
under a $C\!PT$ transformation,
\begin{eqnarray}
\hat{\Theta} \hat{A}^{\dagger} ({\bf x},t) \hat{\Theta}^{-1} =
\eta^{}_{A} \hat{A}(-{\bf x},-t)
\;,
\\
\noalign{\hbox{and}}
\hat{\Theta} \hat{B}^{\dagger} ({\bf x},t) \hat{\Theta}^{-1} =
\eta^{}_{B} \hat{B}(-{\bf x},-t)
\;,
\label{eq:undercpt}
\end{eqnarray}
where $\hat{\Theta}$ is the anti-unitary $C\!PT$ operator and
$|\eta^{}_{A}| = |\eta^{}_{B}| = 1$.
If the equilibrium density matrix
$
\hat{\rho}
\equiv
e^{-\beta\hat{H}}
/
{\rm Tr}\, e^{-\beta\hat{H}}
$
of the theory is $C\!PT$ invariant
({\it i.e.} no chemical potential),
then the following identities holds,
\begin{eqnarray}
\langle [
        \hat{A} ({\bf x},t), \hat{B}(0)
] \rangle ^{\ast}
& \displaystyle =  & \displaystyle
\sum_{n}
\langle \hat{\Theta} n |
\hat{\Theta}
            \hat{\rho} [ \hat{A} ({\bf x},t), \hat{B}(0) ]
\hat{\Theta}^{-1} \hat{\Theta}
| n \rangle
\nonumber\\
& \displaystyle = & \displaystyle
\eta^{}_{A}\eta^{}_{B}
\sum_{n^{\prime}}
\langle n^{\prime} |
                  \hat{\rho} [ \hat{A}^{\dagger} (-{\bf x},-t),
\hat{B}^{\dagger}(0) ]
| n^{\prime} \rangle
\nonumber\\
& \displaystyle  = & \displaystyle
-\eta^{}_{A}\eta^{}_{B}
\sum_{n^{\prime}}
\langle n^{\prime} |
                    \hat{\rho} [ \hat{B}^{\dagger} ({\bf x},t),
\hat{A}^{\dagger} (0) ]
| n^{\prime} \rangle
\nonumber\\
& \displaystyle  = & \displaystyle
-\eta^{}_{A}\eta^{}_{B}
\langle
[ \hat{B}^{\dagger} ({\bf x},t), \hat{A}^{\dagger} (0) ]
\rangle
\;.
\label{eq:cptconsequence}
\end{eqnarray}
Hermiticity implies the following independent relation
\begin{eqnarray}
\langle [
\hat{A} ({\bf x},t), \hat{B}(0)
] \rangle ^{\ast}
& \displaystyle = & \displaystyle
\langle [
\hat{B}^{\dagger} (0) , \hat{A}^{\dagger} ({\bf x},t)
] \rangle
\nonumber\\
& \displaystyle = & \displaystyle
\langle [
\hat{B}^{\dagger} (-{\bf x},-t) , \hat{A}^{\dagger} (0)
] \rangle
\;.
\label{eq:cmplxconjugate}
\end{eqnarray}

In terms of the Fourier transforms,
the relation~(\ref{eq:cptconsequence})
reads
$\rho_{AB}^{\ast}({\bf k}, \omega)
= -\eta^{}_{A}\eta^{}_{B}
    \rho_{B^{\dagger} A^{\dagger}} (-{\bf k},-\omega)$,
and Eq.~(\ref{eq:cmplxconjugate}) gives
$\rho_{AB}^{\ast} ({\bf k}, \omega)
= \rho_{B^{\dagger} A^{\dagger}} ({\bf k}, \omega)$.
Combined, these imply that
\begin{equation}
-\eta^{}_{A}\eta^{}_{B} \,
\rho_{B^{\dagger} A^{\dagger}} (-{\bf k}, -\omega)
=
\rho_{B^{\dagger} A^{\dagger}} ({\bf k}, \omega)
\;.
\label{eq:oddproof}
\end{equation}
If $\hat{B}=\hat{A}^{\dagger}$, then,
the Hermiticity relation~(\ref{eq:cmplxconjugate})
gives
$\rho_{A A^{\dagger}}^{\ast} ({\bf k}, \omega)
= \rho_{A A^{\dagger}} ({\bf k}, \omega)$
implying that
$\rho_{A A^{\dagger}} ({\bf k}, \omega)$ is real.
Since space is isotropic, the spectral density
must also be an even function of the spatial momentum ${\bf k}$.
Thus in any $C\!PT$-invariant equilibrium state the spectral density
$\rho_{A A^{\dagger}} (|{\bf k}|, \omega)$
is an odd function of frequency,
\begin{equation}
\rho_{A A^{\dagger}} (|{\bf k}|, \omega)
= -\rho_{A A^{\dagger}} (|{\bf k}|, -\omega)
\;.
\end{equation}

\appendix{operators with time derivatives}
\label{app:derivatives}

If the external operators involve time derivatives, the sum over the
permutations of the vertices in the upper part
of a cut diagram
is
\begin{eqnarray}
{\cal J}^{+}_{A}( q^0, \{ k^0_{\alpha} \}, \{ k^0_c \} )
& = & \displaystyle
(i)^{n_{+}}
\int_{-\infty}^{\infty}
\prod_{p=n-n_{+}}^n dt_p \,
e^{i q^0 t_l}
\prod_{
{
\hbox{\scriptsize{\rm cut lines}}
\atop
\hbox{\scriptsize{\it c}}
}
}
\!\! e^{-ik^0_c t_d}
\nonumber\\
\; & \; & \displaystyle \qquad {}\times
\hat{O}_{A} ( d/dt_l )
\prod_{
{
\hbox{\scriptsize{\rm uncut lines}}
\atop
\hbox{$\scriptstyle \alpha$}
}
}
\!\! e^{-i(k^0_{\alpha} - i\epsilon) |t_b - t_a|}
\;
\label{eq:appJ+}
\end{eqnarray}
where
$\hat{O}_{A} ( d/dt_l )$
is the action of the external operator $\hat{A}$.
The spatial momentum dependence of $\hat{O}_{A} ( d/dt_l )$ is
suppressed here since it is of no importance in this discussion.
As in the main text,
$t_l$ denotes the vertex where the external frequency
$q^0$ flows out
and all the times
$\{ t_a, t_b, t_d, t_l \}$
are within the set
$\{t_{n-n_{+}}, \cdots, t_n \}$.

If the time-orderings are all reversed,
then the same (now lower) parts of the diagram sum to
\begin{eqnarray}
\; & \vphantom{\int} & \;
\nonumber\\
{\cal J}^{+}_{A}( q^0, \{ k^0_{\alpha} \}, \{ -k^0_c \} )
& = & \displaystyle
(i)^{n_{+}}
\int_{-\infty}^{\infty}
\prod_{p=n-n_{+}}^n dt_p \,
e^{-i q^0 t_l}
\prod_{
{
\hbox{\scriptsize{\rm cut lines}}
\atop
\hbox{\scriptsize{\it c}}
}
}
\!\! e^{-ik^0_c t_d}
\nonumber\\
\; & \; & \displaystyle \qquad {}\times
\hat{O}_{A} ( -d/dt_l )
\prod_{
{
\hbox{\scriptsize{\rm uncut lines}}
\atop
\hbox{$\scriptstyle \alpha$}
}
}
\!\! e^{-i(k^0_{\alpha} - i\epsilon) |t_b - t_a|}
\nonumber\\
\; & = & \displaystyle
(i)^{n_{+}}
\int_{-\infty}^{\infty}
\prod_{p=n-n_{+}}^n dt_p \,
e^{i q^0 t_l}
\prod_{
{
\hbox{\scriptsize{\rm cut lines}}
\atop
\hbox{\scriptsize{\it c}}
}
}
\!\! e^{ik^0_c t_d}
\nonumber\\
\; & \; & \displaystyle \qquad {}\times
\hat{O}_{A} ( d/dt_l )
\prod_{
{
\hbox{\scriptsize{\rm uncut lines}}
\atop
\hbox{$\scriptstyle \alpha$}
}
}
\!\! e^{-i(k^0_{\alpha} - i\epsilon) |t_b - t_a|}
\;.
\end{eqnarray}
In the first equality,
the sign change of $q^0$ to $-q^0$ comes from the change in sign of
the frequency denominators.
The sign change of
the derivative $d/dt_l$ comes from the fact that
factor of frequency brought down by the time derivative must change
sign when the time-ordering is reversed.
Since the factors corresponding
to the uncut lines
$e^{-i(k^0_{\alpha} - i\epsilon) |t_b - t_a|}$
are not changed, the sign of the derivative $d/dt_l$ must change.
The final equality comes from changing the integration
variables $\{ t_p \}$ to $\{ -t_p \}$.

A similar argument holds for the other half of the diagram.
Thus the whole expression for the sum of all cut time-ordered diagrams
with the the external momentum flowing out below the cut is the same
as the expression with external momentum flowing out above the cut
except for changing the sign
of the frequencies of all cut lines. Changing the sign of the
cut frequencies introduces additional exponential factors
coming from the statistical factors.
Using the overall
frequency conservation,
those factors combine to yield $e^{-q^0\beta}$.

\appendix{non-zero chemical potential}
\label{app:chemicalpotential}

Consider a complex scalar field theory
with non-zero chemical potential $\mu$
associated with the conserved charge.
The Euclidean Lagrange density for this theory is given by
\begin{equation}
-{\cal L} =
\phi^{*} ( - \partial_{\tau}^2 - \nabla^2 + m^2 - \mu^2 ) \phi
+ { \textstyle {1 \over 3!} } \lambda (\phi^{*} \phi)^2
+ i\mu
( \phi^* \partial_{\tau} \phi - \phi \partial_{\tau} \phi^* )
\;.
\end{equation}
One may regard the theory as having
two different free propagators,
\begin{eqnarray}
\tilde{G}_{\mu}^E (|{\bf k}|, i\nu)
& \displaystyle \equiv & \displaystyle
\int  d^3 {\bf x} \, \int_0^{\beta} d\tau \,
e^{ -i{\bf k} \cdot {\bf x} + i\nu \tau }
\langle
{\cal T}
(\hat{\phi}^{*} ({\bf x}, -i\tau) \hat{\phi} (0))
\rangle_0
=
{ 1 \over E_k^2 + (\nu - i\mu)^2 }
\\
\noalign{\hbox{and}}
\tilde{G}_{-\mu}^E (|{\bf k}|, i\nu)
& \displaystyle \equiv & \displaystyle
\int d^3 {\bf x} \, \int_0^{\beta} d\tau \,
e^{ -i{\bf k} \cdot {\bf x} + i\nu \tau }
\langle
{\cal T}
(\hat{\phi} ({\bf x}, -i\tau) \hat{\phi}^{*} (0))
\rangle_0
=
{ 1 \over E_k^2 + (\nu + i\mu)^2 }
\;,
\end{eqnarray}
where $\langle \cdots \rangle_0$ indicates
the average is taken with only
the quadratic part of the Lagrangian.
Because of this difference, each line in a Feynman diagram is assigned
a direction which
indicates which end of the line corresponds to $\hat{\phi}^{*}$,
or equivalently
the direction of the flow of conserved charge.
At each interaction vertex,
two lines must point outwards
while the other two point inwards.
In coordinate space, a line pointing to
$( \tau_b, {\bf x}_b )$
from
$( \tau_a, {\bf x}_a )$
corresponds to the propagator
$
\langle
{\cal T}
(\hat{\phi}^{*} ({\bf x}_b , -i\tau_b)
\hat{\phi} ({\bf x}_a , -i\tau_a))
\rangle_0
$,
or in momentum space,
if momentum is
assigned in accordance with the arrow,
then each line represents the propagator
$\tilde{G}_{\mu}^E (|{\bf k}_{\alpha}|, i\nu_{\alpha})$.
There is then no need to introduce
$\tilde{G}_{-\mu}^E (|{\bf k}_{\alpha}|, i\nu_{\alpha})$.

The $C\!PT$ invariance of the system is explicitly broken by the
chemical potential term.
Nevertheless, an argument similar to the one
given in Appendix~\ref{app:cpt} can be applied
using the relation
\begin{equation}
\hat{\Theta}
\hat{\rho}
\hat{\Theta}^{-1}
=
\hat{\Theta}
{
e^{-\beta\hat{H} + \mu \hat{Q}}
\over
{\rm Tr} e^{-\beta\hat{H} + \mu \hat{Q}}
}
\hat{\Theta}^{-1}
=
{
e^{-\beta\hat{H} - \mu \hat{Q}}
\over
{\rm Tr} e^{-\beta\hat{H} - \mu \hat{Q}}
}
\;.
\label{eq:rho}
\end{equation}
The one-particle spectral density
is not an odd function of frequency
but satisfies
\begin{eqnarray}
\rho_{\mu} (|{\bf k}|,\omega)
& \displaystyle \equiv & \displaystyle
{-i} \,
{\rm Disc}\,
\tilde{G}_{\mu} (|{\bf k}|, \omega)
\nonumber\\
\;
& \displaystyle = & \displaystyle
-\rho_{-\mu} (|{\bf k}|, -\omega)
\;.
\label{eq:sdsymmetry}
\end{eqnarray}
Here $\tilde{G}_{\mu} (|{\bf k}|, i\nu)$ is the fully dressed
single-particle
propagator in the imaginary-time formalism.
The ``mixed'' propagator is then given by
\begin{eqnarray}
\displaystyle
\tilde{G}_{\mu} (|{\bf k}|, \tau)
=
\int_{-\infty}^{\infty} {d\omega \over 2\pi}\,
\Bigl(
\rho_{\mu} (|{\bf k}|,\omega)  \,   \theta (\tau) +
\rho_{-\mu} (|{\bf k}|,\omega)  \,   \theta (-\tau)
\Bigr) \,
e^{-\omega|\tau|} \,
\Bigl( 1+n(\omega) \Bigr)
\;.
\label{eq:mixedwithchemicalagain}
\end{eqnarray}

As in the main text, consider a Feynman diagram with a total of
$n{+}1$ vertices.
Using the mixed propagators,
the contribution of a line with a momentum ${\bf k}_{\alpha}$
pointing to a time $\tau_b$ from time $\tau_a$ is
$\tilde{G}_{\mu} (|{\bf k}_{\alpha}|, \tau_b {-} \tau_a)$.
To evaluate the resulting time integral,
one of the times is again set to zero,
and the integrand is split into
$n!$ terms according to the relative time orderings of remaining $n$
time variables.
A typical time integral still is
\begin{equation}
{\cal I} (\{ k^0_{\alpha} \}, i\nu)
\equiv
\int_{0}^{\beta} d\tau_{n} \cdots \int_{0}^{\tau_3} d\tau_2
\int_{0}^{\tau_2} d\tau_1 \,
\exp \Bigl\{
\sum_{l=1}^{n} \,
-
\sigma_l \tau_l
\Bigr\}
\;,
\label{eq:apponetypicalterm}
\end{equation}
as in Eq. (\ref{eq:onetypicalterm}).
Which combination of
$\rho_{\mu} (|{\bf k}_{\alpha}|, k^0_{\alpha})$ and
$\rho_{-\mu} (|{\bf k}_{\alpha}|, k^0_{\alpha})$
multiplies this time integral
depends on the particular time ordering that gives
rise to the integral.
To determine this, draw the corresponding time-ordered diagram
with the directions of the arrows
determined by the flow of charge.
If the arrow of a line $\alpha$
points upwards, it corresponds to a factor of
$(1+n(k^0_{\alpha}))\,
\rho_{\mu} (|{\bf k}_{\alpha}|, k^0_{\alpha})$,
and if it points downwards,
then it corresponds to
$(1+n(k^0_{\alpha}))\,
\rho_{-\mu} (|{\bf k}_{\alpha}|, k^0_{\alpha})$.

The time integrations of
Eq.~(\ref{eq:apponetypicalterm}) produce
\begin{equation}
{\cal I} (\{ k^0_{\alpha} \}, i\nu)
=
\lim_{\epsilon \to 0}
\,
\sum_{k=0}^n
       \left(
             \prod_{j=1}^k (\Omega_{j}^k - i\epsilon_j^k) \,
                    \cdot \,
        \prod_{j'=k+1}^n (-\Omega_{k+1}^{j'} + i\epsilon_{k+1}^{j'})
       \right)^{-1}
    \exp \left\{
       \beta (-\Omega_{k+1}^n + i\epsilon_{k+1}^n)
         \right\}
\;,
\end{equation}
with the same notation used in the main text.
Once again,
the factor $e^{-\beta {\rm Re}\, \Omega_{k+1}^n}$ may
be omitted if, at the same
time, the sign of all frequencies in the set
$\Lambda_{\rm in}^{k+1, n}$,
and the sign of the chemical potentials
associated with the lines in this set, are also changed.
The resulting factors of $\Lambda_j^k$ have the same
interpretation as the sum of the frequencies of all lines crossing
the interval between $\tau_j$ and $\tau_{j-1}$
when the time ordering is
$\tau_k \geq \tau_{k-1} \geq \cdots \geq \tau_j \geq \cdots \geq 0
\geq \tau_n \geq \tau_{n-1} \geq \cdots \geq \tau_{k+1}$.
This time ordering is achieved by moving the sequence
of times
$\tau_n \geq \tau_{n-1} \geq \cdots \geq \tau_{k+1}$ down below zero.
Diagrammatically, the only arrows that change directions under this
operation are those of the lines in the set
$\Lambda_{\rm in}^{k+1, n}$, or equivalently the lines whose
chemical potentials had their sign flipped.
Thus, it remains true
that every upward arrow is assigned to a factor
$\rho_{\mu}$ and a downward arrow to
every factor of $\rho_{-\mu}$.
As before, the contribution of
all time-ordered diagrams is expressed
in terms of statistical factors, spectral densities
and energy denominators.
The process of extracting the spectral density is not affected by the
presence of the non-zero chemical potential.
The previous discussion
in the main text applys without change.
The resummation of the cut time-ordered
diagrams is possible in
terms of a real-time propagator,
\begin{eqnarray}
\tilde{G}_{\mu} (k)
& \equiv & \displaystyle
\int dt\,
e^{ik^0t}
\int {d\omega \over 2\pi}\,
\Bigl(
\rho_{\mu} (|{\bf k}|,\omega)  \,   \theta (\tau) +
\rho_{-\mu} (|{\bf k}|,\omega)  \,   \theta (-\tau)
\Bigr) \,
e^{-i\omega|t|} \,
\Bigl( 1+n(\omega) \Bigr)
\nonumber\\
\; & = & \displaystyle
\int {d\omega \over 2\pi i} \,
\Bigl( 1 + n(\omega) \Bigr)
\left(
{
 \rho_{\mu} (|{\bf k}|,\omega)
        \over
\omega - k^0 - i \epsilon
}
+
{
 \rho_{-\mu} (|{\bf k}|,\omega)
        \over
\omega + k^0 - i \epsilon
}
\right) \,
\;.
\end{eqnarray}
The resulting diagrammatic rules are:
\begin{enumerate}
\item Draw all suitable topologically distinct cut diagrams
with directed lines.

\item Label each line with a 4-momentum $k_{\alpha}$ flowing in the
direction of the arrow.
Assign external 4-momentum where
the operators are inserted.
Conserve 4-momentum at each vertex.

\item On the unshaded side of the cut, use the standard Feynman rules;
assign
$
\tilde{G}_{\mu} (k_{\alpha})
$
to each uncut line $\alpha$,
and assign $-i\lambda$ to each interaction vertex.

\item On the shaded side, use the complex conjugated Feynman rules;
assign
$ \tilde{G}_{\mu} (k_{\alpha})^* $
to each uncut line $\alpha$,
and assign $i\lambda$ to each interaction vertex.

\item To each cut line
which points towards the unshaded region,
assign
$$
\tilde{\Delta}_{+} (k)
\equiv
\Bigl( 1+n(k^0) \Bigr)
\rho_{\mu} (|{\bf k}|, k^0)
\;.
$$
To each cut line
which points towards the shaded region,
assign
$$
\tilde{\Delta}_{-} (k)
\equiv
n(k^0)
\rho_{-\mu} (|{\bf k}|, k^0)
\;.
$$

\item Integrate over all loop momenta, and divide by the symmetry
factor. If the external frequency flows out of the shaded region,
assign an additional factor of -1.
\end{enumerate}
\end{document}